\documentclass[preprint2]{aastex6}

\fullcollaborationName{The Friends of AASTeX Collaboration}

\shorttitle{Star-forming regions beyond the outer Arm}
\shortauthors{Izumi et al.}

\begin{document}


\title{Star formation activity beyond the outer Arm I: {\it WISE}-selected candidate star-forming regions}


\author{Natsuko Izumi\altaffilmark{1,2}, Naoto Kobayashi\altaffilmark{2,3,4}, Chikako Yasui\altaffilmark{1,2},
Masao Saito\altaffilmark{1,5}, and Satoshi Hamano\altaffilmark{2}}
 
\altaffiltext{1}{National Astronomical Observatory of Japan, 2-21-1, Osawa, Mitaka, Tokyo, 181-8588, Japan, email: {\tt natsuko.izumi@nao.ac.jp}}
\altaffiltext{2}{Laboratory of Infrared High-resolution spectroscopy (LIH), Koyama Astronomical Observatory, Kyoto Sangyo University, Motoyama, Kamigamo, Kita-ku, Kyoto 603-8555, Japan}
\altaffiltext{3}{Institute of Astronomy, School of Science, University of Tokyo, 2-21-1, Osawa, Mitaka, Tokyo 181-0015, Japan}
\altaffiltext{4}{Kiso Observatory, Institute of Astronomy, School of Science, University of Tokyo,10762-30 Mitake, Kiso-machi, Kiso-gun, Nagano 397-0101, Japan}
\altaffiltext{5}{The Graduate University of Advanced Studies, (SOKENDAI), 2-21-1 Osawa, Mitaka, Tokyo 181-8588, Japan}

\begin{abstract}
The outer Galaxy beyond the Outer Arm provides a good opportunity to study star formation in an environment significantly different from that in the solar neighborhood.
However, star-forming regions in the outer Galaxy have never been comprehensively studied or cataloged because of the difficulties in detecting them at such large distances.
We studied 33 known young star-forming regions associated with 13 molecular clouds at  $R_{\rm G}$ $\ge$ 13.5 kpc in the outer Galaxy
with data from the {\em Wide-field Infrared Survey Explorer} ({\em WISE}) mid-infrared all-sky survey.
From their color distribution, we developed a simple identification criterion of star-forming regions in the outer Galaxy with the {\em WISE} color.
We applied the criterion to all the {\em WISE} sources in the molecular clouds in the outer Galaxy at $R_{\rm G}$ $\ge$ 13.5 kpc
detected with the Five College Radio Astronomy Observatory (FCRAO) $^{12}$CO survey of the outer Galaxy,
of which the survey region is 102$^\circ$.49 $\le$ $l$ $\le$ 141$^\circ$.54, $-$3$^\circ$.03 $\le$ $b$ $\le$ 5$^\circ$.41,
and successfully identified 711 new candidate star-forming regions in 240 molecular clouds.
The large number of samples enables us to perform the statistical study of star-formation properties in the outer Galaxy for the first time.
This study is crucial to investigate the fundamental star-formation properties, including star-formation rate, star-formation efficiency, and initial mass function,
in a primordial environment such as the early phase of the Galaxy formation.
\end{abstract}

\section{Introduction} \label{sec:1}
In the past, star formation has been studied in detail in nearby star-forming regions ($D$ $<$ 1 kpc), such as Orion or Taurus.
As a result, star formation properties, including star-formation rate (SFR), star-formation efficiency (SFE), and stellar initial mass function (IMF),
have been derived accurately in each star-forming region \citep[e.g.][]{Bastian10,Kennicutt12}.
The important question is to address whether the same properties hold true even in a primordial environment with low gas density and low metallicity,
such as in the early phase of the formation of the Galaxy.

In low-gas-density environments, the SFR is known to decrease significantly
from the empirical relation between SFR surface density ($\Sigma_{\rm SFR}$) and total (atomic and molecular) gas surface density ($\Sigma _{\rm gas}$)
known as the Kennicutt-Schmidt law \citep[$\Sigma_{\rm SFR}$ $\propto$ $\Sigma _{\rm gas}^n$;][]{Schmidt59,Kennicutt98,Kennicutt12}:
the exponent $n$ in the Kennicutt-Schmidt law increases from the usual $n$ $\sim$ 1 (10 $M_\odot$pc$^{-2}$ $\lesssim$ $\Sigma _{\rm gas}$ $\lesssim$ 100 $M_\odot$pc$^{-2}$) 
to $n$ $\sim$ 2 in regions that have surface density less than 10 $M_\odot$pc$^{-2}$ \citep[e.g. Figure 15 in ][]{Bigiel08}.
The recent {\em Herschel} observations of nearby dwarf galaxies suggest that constant SFE (cSFE; $\Sigma_{\rm SFR}$/$\Sigma _{\rm gas}$)
decreases significantly in low-metallicity environments \citep{Shi14}.
However, the mechanisms for this qualitative change of star formation have not been understood, because detailed observations of star-forming regions have been impossible even in nearby galaxies, not to mention in high-{\em z} galaxies.

The outer Galaxy beyond the Outer Arm has an environment significantly different from the solar neighborhood, with a much lower gas density \citep{Wolfire03} and lower metallicity \citep{Smartt97}.
The region is often called the far outer Galaxy (FOG), which is nominally defined as the region with a galactocentric radius of more than 13.5 kpc \citep[e.g.][]{Snell02}.
Furthermore, we have defined the extreme outer Galaxy (EOG) for the region of $R_{\rm G}$ $\ge$ 18 kpc \citep{Kobayashi08,Yasui08}, where little or no perturbation from the spiral arm is expected.
The EOG serves as an excellent laboratory for studying the star-forming processes in an environment that has characteristics similar to those in our Galaxy in the early phase of the formation 
and dwarf galaxies \citep{Ferguson98, Kobayashi08}.
In our Galaxy, the SFR and cSFE have been derived up to $R_{\rm G}$ $\sim$ 15 kpc \citep{Kennicutt12},
and they clearly show those trends;
the SFR and cSFE start to decrease significantly at $R_{\rm G}$ $\sim$ 13.5 kpc
and then drop at $R_{\rm G}$ $\sim$ 15 kpc to roughly one-eighth and one-fourth of those in the solar neighborhood, respectively.
The FOG and EOG are much closer than any galaxies; hence, the most detailed study of the star-forming region is well possible, as presented in this paper.

\begin{figure*}
\epsscale{0.8}
\plotone{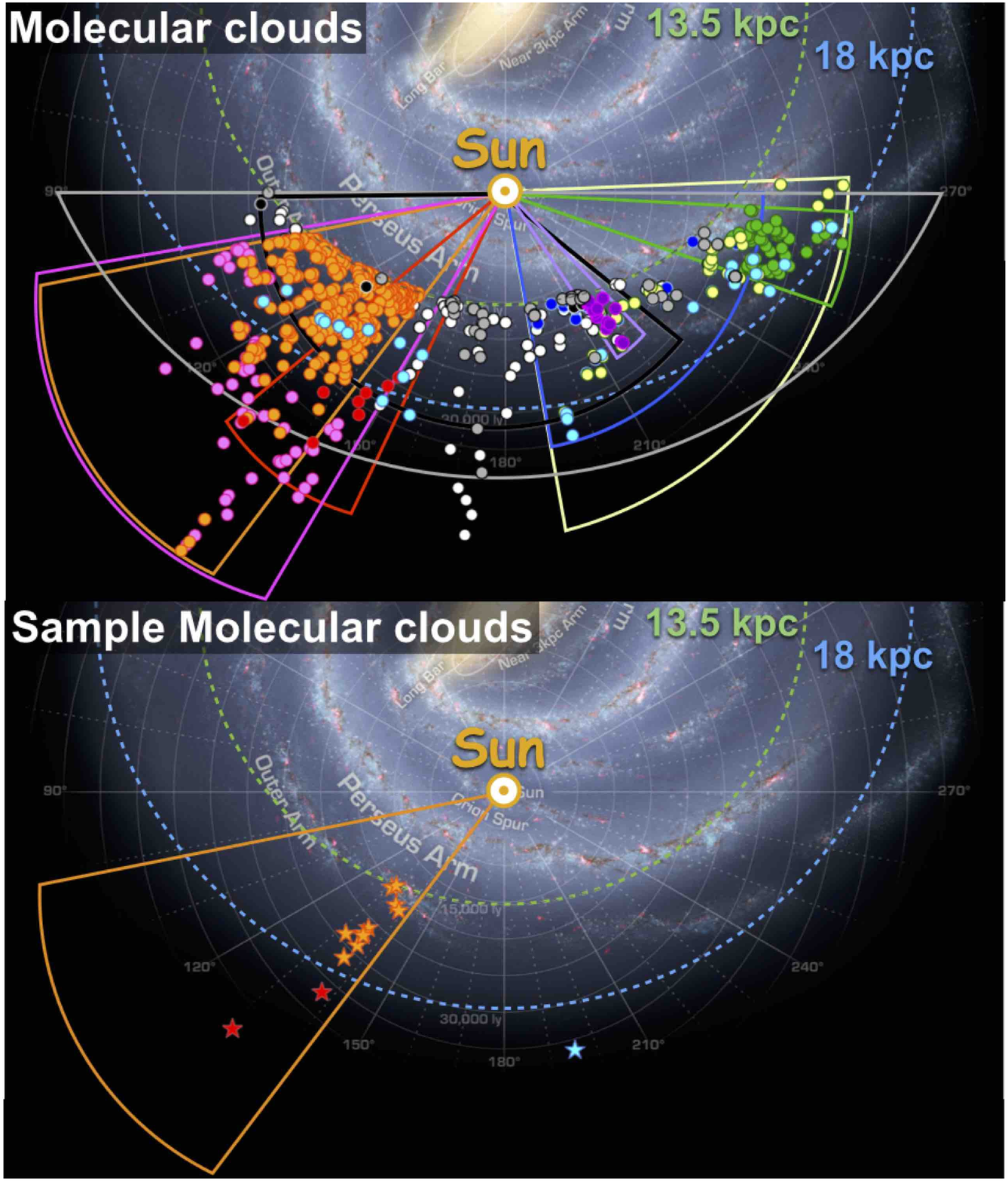}
\caption{
Top: distribution of molecular clouds discovered by representative surveys at the second and third quadrants
(orange: \cite{Brunt03}; magenta: \citet{Sun15};
red:  \citet{Digel94,Smartt96,Kobayashi08}; cyan: \citet{Brand94}; blue: \citet{May97}; green: \citet{Nakagawa05}; yellow: \citet{May93,Vazquez08};
black: \citet{Kutner83,Mead88}; white: \citet{Yang02}; purple: \citet{Elia13}; gray: \citet{Dame01,Rice16}). 
Fan-shaped regions show each survey area, corresponding to the circles of the same color
(orange fan shape indicates the survey area of the FCRAO CO survey of the outer Galaxy: 102$^\circ$.49 $\le$ $l$ $\le$ 141$^\circ$.54, $-$3$^\circ$.03 $\le$ $b$ $\le$ 5$^\circ$.41,
which is also that of the survey of star-forming regions presented in this paper).
Note that \citet{Brand94} and \citet{Yang02} did not set a survey area but rather searched for molecular clouds based on {\it IRAS} sources in the second and third quadrants. 
\citet{Kutner83} and \citet{Mead88} also searched molecular clouds in the first quadrant, and 
\citet{Dame01} and \citet{Rice16} searched molecular clouds in the entire Galactic plane; therefore, we picked up the discovered molecular clouds and survey area
only in the second and third quadrants.
We derived the kinematic distance (and galactocentric radius) of all clouds in the figure assuming that the rotation speed of the Sun and clouds is 220 km s$^{-1}$ and that the galactocentric distance of the Sun is 8.5 kpc.
Bottom: distribution of our sample distant molecular clouds with associated star-forming regions beyond the Outer Arm on the Galactic plane.
Red and cyan star symbols show clouds in the EOG (red: Digel Clouds 1 and 2 \citep[e.g.][]{Digel94}, cyan: WB89-789 cloud \citep[e.g.][]{Brand07}.
Orange star symbols show clouds in the FOG \citep[e.g.][]{Snell02}.
The orange fan shape is the same as the survey area in the top panel.
The locations of those clouds are tabulated in Table \ref{tbl:sampleMClist}.
}
\label{MC_dist} 
\end{figure*}

Several hundreds ($\sim$ 500) of molecular clouds
have been detected in the FOG \citep[e.g.][]{Snell02,Brunt03} and EOG \citep[e.g.][]{Digel94} up to $R_{\rm G}$ $\sim$ 22 kpc (top panel of Figure \ref{MC_dist}).
Compared with the solar neighborhood, however, many fewer star-forming regions have been identified \citep[e.g.][]{Yasui06,Brand07,Izumi14},
primarily because of the difficulties in detecting them at such large distances, let alone determining their distances.
These kind of studies have been traditionally conducted with the {\em Infrared Astronomical Satellite} \citep[{\em IRAS};][]{Neugebauer84,Beichman88}
infrared (IR) all-sky survey data \citep[e.g.][]{Hughes89,Kerton03}.
The recent mid-infrared (MIR) all-sky survey explorer {\em Wide-field Infrared Survey Explorer} \citep[{\em WISE};][]{Wright10,Jarrett11}
has achieved a major increase in sensitivity: about 100 times greater than {\it IRAS} \citep{Wright10}.
Therefore, the {\em WISE} data have a great potential in searching for distant star-forming regions.
In the past studies, {\em WISE} magnitude and color for individual young stellar objects (YSOs) in the solar neighborhood ($D$ $\le$ 2 kpc)
have been established \citep[e.g.][]{Koenig14} by comparing them with the well-established data from the {\em Spitzer Space Telescope} \citep{Werner04}.

Considering the typical distance between YSOs in young clusters (2$^\prime$ -- 3$^\prime$ for the Taurus star-forming region at $D$ $\sim$ 150 pc),
we cannot resolve clusters beyond $D$ $\sim$ 4 kpc into individual stars with the resolution of {\it WISE} ($\sim$6$^{\prime \prime}$ -- 12$^{\prime \prime}$).
In this paper, we study the {\it WISE} magnitudes and colors of known young star-forming regions, without resolving them, in the FOG/EOG
and aim to develop a simple identification criterion to identify unresolved star-forming regions at a large distance.
Furthermore, we select CO molecular clouds from a number of surveys as the reference to our study, given that our targets,
young star-forming regions (age $<$ 3 Myr), should accompany their parental molecular clouds \citep{Lada03}.
We will then apply the criterion to all the {\it WISE} data in CO molecular clouds detected with the 
Five College Radio Observatory (FCRAO) $^{12}$CO survey of the outer Galaxy \citep{Heyer98},
which has achieved the best combination of large-areacoverage and high sensitivity among all the available outer Galaxy surveys.

The goal of our study is to understand the global properties of star-formation activities beyond the Outer Arm in the scale of molecular clouds.
In this paper, we describe our method of searching for new star-forming regions in the FOG and EOG with the {\em WISE} data
and present the result. 
Subsequent papers will focus on the distribution of newly identified star-forming regions beyond the Outer Arm, the properties of those star-forming regions,
and the properties of molecular clouds with and without associated star-forming regions.

\section{Data} \label{sec:2}
In this paper, we use the {\it WISE} MIR all-sky survey data \citep{Wright10,Jarrett11} and FCRAO $^{12}$CO survey data of the outer Galaxy \citep{Heyer98}.

\subsection{WISE All-sky Data} \label{sec:2.1}
{\it WISE} mapped at least eight times over 99 \% of the sky in four MIR bands centered at 3.4, 4.6, 12, and 22 {\micron} in the six-month survey in 2010 \citep{Wright10}.
{\em WISE} achieved 5$\sigma$ detection thresholds for a pointlike source of 16.5, 15.5, 11.2, and 7.9 mag (Vega magnitude) at the 3.4, 4.6, 12, and 22 $\micron$ bands, respectively,
in regions observed eight times or more \citep{Wright10}.
The angular resolutions are 6$^{\prime \prime}$.1, 6$^{\prime \prime}$.4, 6$^{\prime \prime}$.5, and 12$^{\prime \prime}$.0, respectively, for the four bands \citep{Wright10}.
The standard spectral energy distribution (SED) model of galaxies \citep[e.g.][]{Cunha08} implies that the 3.4 and 4.6 {\micron} bands show emission mainly from stars,
while the 12 and 22 {\micron} bands show emission mainly from circumstellar dust.

We used the AllWISE Source Catalog\footnote{\url{http://wise2.ipac.caltech.edu/docs/release/allwise/}}
to obtain the {\it WISE} magnitudes and colors of the sample star-forming regions.
The AllWISE Source Catalog contains astrometry and photometry for 747,634,026 objects detected on the deep
AllWISE Atlas Intensity Images\footnote{Detailed information is given at \url{http://wise2.ipac.caltech.edu/docs/release/allwise/expsup/}}.
In addition to the cataloged positions and photometric information, we also used the measurement quality and source reliability information,
especially the contamination and confusion flags ({\it cc\_flags}), to reject false sources.
The {\it cc\_flags} show that the source may be either spurious due to a diffraction spike (D), short-term latent image (P), scattered-light halo (H), or optical ghost image (O).
In this paper, we normally used only the {\it WISE} sources without {\it cc\_flags} in all of the four bands 
and with a signal-to-noise ratio (S/N) of higher than 5 in all of the 3.4, 4.6, and 12 {\micron} bands.
Those three bands are crucial for source classification of YSOs in the solar neighborhood \citep[e.g. Figure 10 in][]{Koenig14}.
We did not use the 22 {\micron} data in our criterion, because the {\it WISE} spatial resolution and sensitivity at this band are much lower than those of the others.

\subsection{FCRAO CO survey data} \label{sec:2.2}
The FCRAO $^{12}$CO survey of the outer Galaxy \citep{Heyer98} has achieved the best combination of
large-area coverage and high sensitivity among all of the available CO surveys of the outer Galaxy.
This survey covers 102$^\circ$.49 $\le$ $l$ $\le$ 141$^\circ$.54 and $-$3$^\circ$.03 $\le$ $b$ $\le$ 5$^\circ$.41 with a total area of about 320 deg$^2$ \citep{Heyer98}.
The median main-beam sensitivity (1 $\sigma$) per channel is $\sim$ 0.9 K with a spatial resolution of  45$^{\prime \prime}$ \citep{Heyer98, Heyer01}.
The $v_{\rm LSR}$ range covers $-$153 km s$^{-1}$ $\le$ $v_{\rm LSR}$ $\le$ +40 km s$^{-1}$ with a velocity resolution of 0.98 km s$^{-1}$ \citep{Heyer98}.
\citet{Heyer01} identified 10,156 clouds in the whole parameter space of the survey.
Later, \citet{Brunt00} reprocessed the data to remove correlated noise induced by reference sharing and contaminating emission present in the reference positions.
They also convolved the data to the spatial resolution of 100$^{\prime \prime}$.44 to be incorporated into the Canadian Galactic Plane Survey \citep{Taylor03}.
As a result, the typical sensitivity of the data was improved to be 0.17 K.
With the better sensitivity achieved, \citet{Brunt03} then identified 14,592 clouds in the whole survey area, which are almost 50 \% more than those of the original survey
despite the smaller velocity range of $-$120 km s$^{-1}$ $\le$ $v_{\rm LSR}$ $\le$ +20.8 km s$^{-1}$ that they used.

In this paper, we employed the molecular cloud catalog by \citet{Brunt03} (hereafter the BKP catalog) to make use of the larger number of clouds compared with the original catalog by \citet{Heyer01}.
The high-sensitivity BKP catalog is essential to study distant star-forming regions in the outer Galaxy.
Using the Galactic ($l$,$b$) coordinates and the $v_{\rm LSR}$ of the clouds in the BKP catalog (hereafter, BKP clouds),
we derived the kinematic distance of BKP clouds, assuming that the rotation speed of the Sun and BKP clouds is 220 km s$^{-1}$
and that the Galactocentric distance of the Sun is 8.5 kpc, and picked up 466 clouds in the outer Galaxy ($R_{\rm G}$ $\ge$ 13.5 kpc) out of 14,592 clouds.
We also estimated the masses of BKP clouds from the integrated CO line intensity ($\int T_{\rm B} dv$), assuming the Galactic average mass-calibration ratio N(H$_2$)/$\int T_{\rm B} dv$
of 2.0 × 10$^{20}$ cm$^{-2}$ (K km s$^{-1}$)$^{-1}$ \citep[e.g.][]{Bolatto13} and the correction for the abundance of helium \citep[1.36;][]{Kennicutt12}.

\section{Identification criteria of distant star-forming regions}  \label{sec:3}
In this section, we investigate the {\it WISE} magnitudes and colors of sample star-forming regions in the outer Galaxy beyond the Outer Arm
and develop an identification criterion for star-forming regions, which will then be applied to unresolved (distant) sources to select candidate star forming regions
in the later section.

\subsection{Sample star-forming regions} \label{sec:3.1}
For the sample star-forming regions, we selected 13 known distant molecular clouds with associated star-forming regions
in the outer Galaxy that have been established by near-infrared (NIR) imaging observations:
three in the EOG and 10 in the FOG.
Table \ref{tbl:sampleMClist} lists them, and the bottom panel of Figure \ref{MC_dist} plots their locations in the Galactic plane.

The three molecular clouds in the EOG are Digel Cloud~1 \citep[$R_{\rm G}$ = 22 kpc;][]{Digel94,Izumi14},
Digel Cloud~2 \citep[$R_{\rm G}$ = 19 kpc;][]{Kobayashi00,Yasui06,Kobayashi08,Yasui08}, and 
the cloud associated WB89-789 (IRAS 0615+1455) \citep[$R_{\rm G}$ = 20.2 kpc;][hereafter the WB89-789 cloud]{Brand07}.
We analyzed the multiwavelength data of Digel Clouds~1 and 2, including high-resolution $^{12}$CO maps with the Nobeyama 45 m radio telescope \citep{Izumi14},
deep NIR ($J,\, H,\, K_S$) images of embedded clusters with Multi-Object Infrared Camera (MOIRCS) on the Subaru 8.2 m telescope \citep{Yasui08,Izumi14},
and wide-field NIR images of Cloud~2 with QUick InfRared Camera (QUIRC) on the University of Hawaii 2.2 m telescope \citep{Kobayashi08}.
We then identified, in addition to the eight star-forming regions (clusters/stellar aggregates) already listed in our previous papers \citep[e.g.][]{Kobayashi00,Kobayashi08,Izumi14},
13 new stellar aggregates (two in Cloud~1 and 11 in Cloud~2) in and near the CO peaks or ridges as reddened stellar associations in the NIR images
(indicated by ``Q'' in the third column of Table \ref{tbl:sampleSFR}; see also Figure \ref{NIR-DC12SFR} and N. Izumi et al. 2017, in preparation).
The other cloud in the EOG, the WB89-789 cloud, was identified by \citet{Brand07}.
To study stars and molecular gas in the direction of WB89-789, they performed multiwavelength observations:
spectroscopy at about 3200--4900 {\AA} with the Telescopio Nazionale Galileo,
NIR ($J,\, H,\, K$) imaging with the European Southern Observatory (ESO) 2.2 m telescope,
molecular lines
($^{12}$CO(2--1), $^{12}$CO(3--2), $^{13}$CO(2--1), C$^{18}$O(3--2) with 15 m James Clerk Maxwell Telescope (JCMT),
and CS(2--1), CS(3--2), and CS(5--4) with the Institute for Radio Astronomy in the Millimetre Range (IRAM) 30 m telescope),
and milli-meter continuum (450 and 850 {\micron} with JCMT, and 1.2 mm with the 15 m Swedish-ESO Submillimetre Telescope).
As a result, they detected a cluster of about 60 stars in the vicinity of the CO peak \citep[Figures 1, 3, and 6 in][]{Brand07}.

In the FOG, the number of reported molecular clouds with associatied star-forming regions has been gradually increasing \citep[e.g.][]{Brand94,Snell02,Yun15}.
Among them, we selected 10 molecular clouds identified by \citet{Snell02}
because they present a relatively large number of samples and performed systematic studies.
Their targets are 10 {\it IRAS} point sources associated with 
10 molecular clouds that had been detected with the FCRAO CO survey of the outer Galaxy \citep{Heyer98}.
They performed $K^\prime$- band imaging of those targets with  QUIRC on the University of Hawaii 2.2 m telescope,
and detected 11 clusters in the vicinity of the CO peaks.
Among them, three were found to be associated with \ion{H}{2} regions: 02071+6235, 02383+6241a, and 02421+6233 (see Table \ref{tbl:sampleSFR}), 
which must be produced by OB stars \citep{Rudolph96,Snell02}.

\subsection{WISE data of sample star-forming regions} \label{sec:3.2}
\subsubsection{WISE images} \label{sec:3.2.1}
\begin{figure*}
\epsscale{1.0}
\plotone{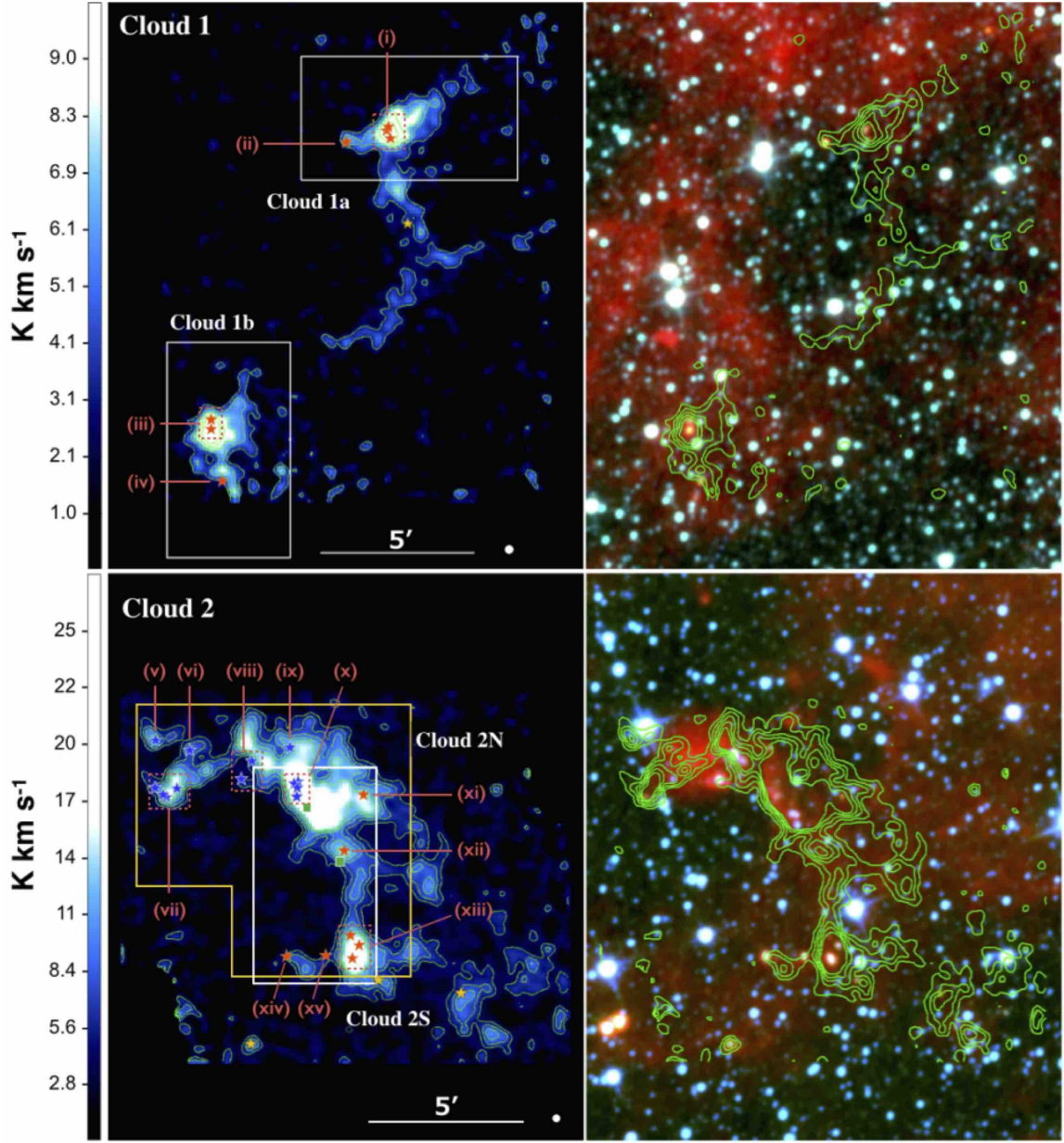}
\caption{
Left: $^{12}$CO(1-0) distribution of the four CO peaks in the Clouds 1 and 2
(with our NRO 45 m telescope data; Cloud 1: $v_{\rm LSR}$ = $-$105.4 to $-$98.9 km s$^{-1}$; Cloud 2: $v_{\rm LSR}$ = $-$106.1 to $-$99.1 km s$^{-1}$).
Green contours show the $^{12}$CO(1-0) distribution with contour levels of 3$\sigma$, 5$\sigma$, 7$\sigma$, 9$\sigma$, and 11$\sigma$
(Cloud 1: 1$\sigma$ = 0.85 K km s$^{-1}$, Cloud 2: 1$\sigma$ = 1.2 K km s$^{-1}$).
Filled circles in the bottom right corners show the resolution of the NRO 45 m telescope ($\sim$ 17$^{\prime \prime}$).
The red star symbols show the locations of star-forming regions (clusters/stellar aggregates) confirmed with our deep NIR images 
\citep[][Section \ref{sec:3.1} in this paper; N. Izumi et al. 2017, in preparation]{Kobayashi08, Yasui08}.
The blue star symbols show the locations of the confirmed star-forming regions with P flags (see Section \ref{sec:2.1} and \ref{sec:3.2.2}).
The yellow star symbols show the locations of the new candidate star-forming regions identified with the {\it WISE} data (see Section \ref{sec:3.4}).
Green squares show the locations of contamination by possible AGNs (see Section \ref{sec:3.5}).
The white boxes and a yellow polygon show the fields of view of the Subaru MOIRCS (4$^\prime$ $\times$ 7$^\prime$) and UH QUIRC (3$^\prime$.2 $\times$ 3$^\prime$.2), respectively.
The red characters and dotted boxes indicate the positions of the star-forming regions for which Figure \ref{NIR-DC12SFR}
gives the zoomed images.
Right: {\it WISE} 3.4 (blue), 4.6 (green), and 12 {\micron} (red) pseudo-color images of the same area. 
Green contours are the same as the $^{12}$CO map in the corresponding left panel. 
}  
\label{CO-NIR-MIR-DC12} 
\end{figure*}

\begin{figure*}
\epsscale{0.8}
\plotone{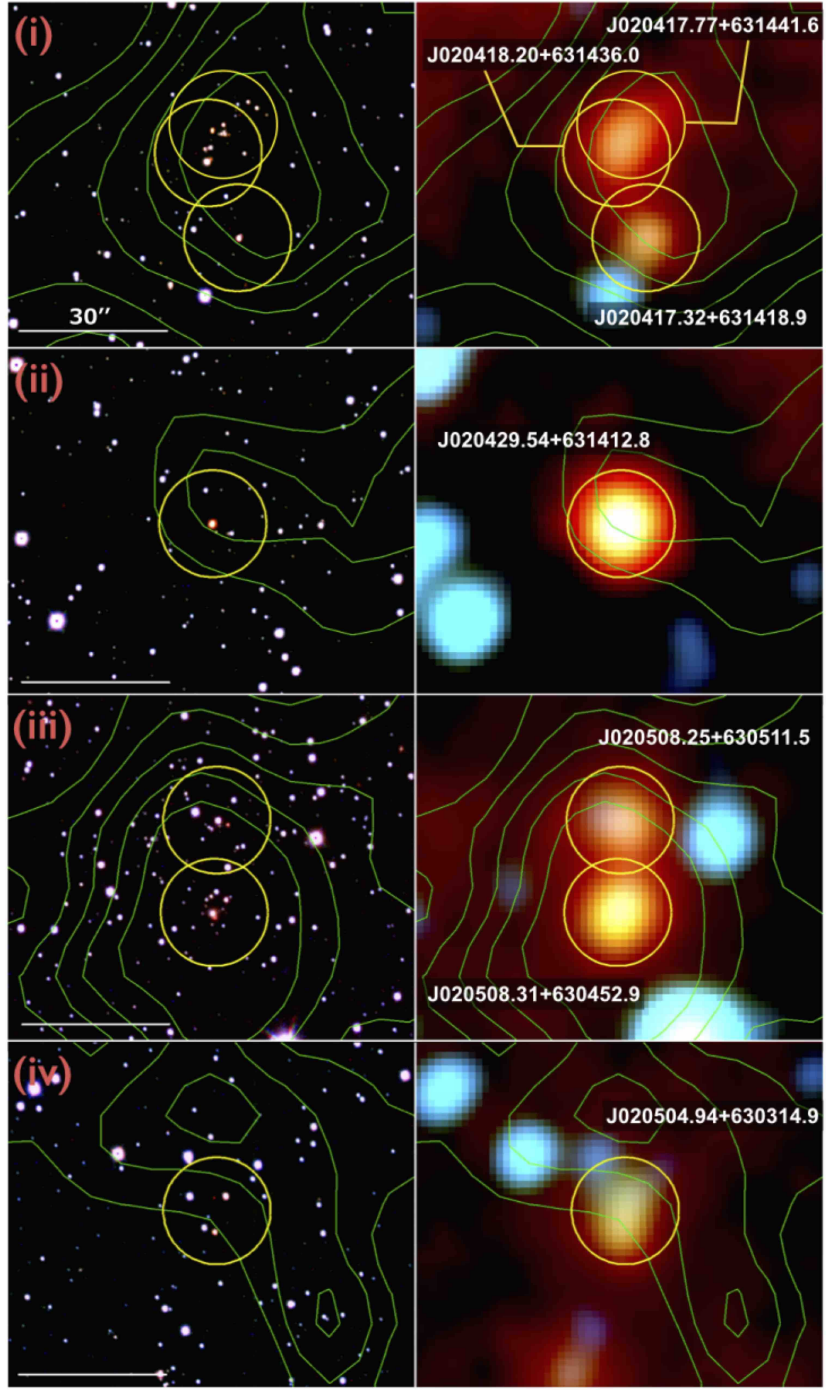}
\caption{
Left: $JHK_{S}$ pseudo-color images taken with Subaru (panels (i)--(iv), (x)--(xiv)) and QUIRC ((v)--(ix)),
of known star-forming regions in the Digel Clouds 1 and 2 (see Figure \ref{CO-NIR-MIR-DC12} for the images of the whole area).
Green contours show the $^{12}$CO(1-0) distribution from the NRO 45 m telescope with contour levels of 3$\sigma$, 5$\sigma$, 7$\sigma$, 9$\sigma$, and 11$\sigma$.
Yellow circles show the location of star-forming regions.
Right: {\it WISE} 3.4 (blue), 4.6 (green), and 12 {\micron} (red) pseudo-color images of the same area. 
Green contours and yellow circles are the same as those in the corresponding left panel.
}  
\label{NIR-DC12SFR} 
\end{figure*}

To investigate unresolved star-forming regions with the {\it WISE} data, we first checked the distribution of CO and the {\it WISE} MIR images of  molecular clouds
with associated sample star-forming regions.
Figure \ref{CO-NIR-MIR-DC12} shows the images of Clouds~1 and 2 in CO (left panels) and {\it WISE} color (constructed from the 3.4, 4.6, and 12 {\micron} bands; right panels).
The red and blue star symbols in Figure \ref{CO-NIR-MIR-DC12} show the locations of the sample star-forming regions.
Figure \ref{NIR-DC12SFR} shows the zoomed-in images of each star-forming region in the clouds
(left panels: Subaru and QUIRC {\it J, H, {\rm and} Ks} bands, right panels: {\it WISE} 3.4, 4.6, and 12 {\micron} bands).
All of the sample star-forming regions, even the very faint ones (stellar aggregates), were clearly detected as reddened sources in the {\it WISE} images.
As expected none of the sample star-forming regions were resolved into individual stars with the resolution of the {\it WISE}.
All of the sample star-forming regions but one were detected in the {\it WISE} images as compact sources.
The sole exception is the Cloud~2-N cluster (region x in Figure \ref{CO-NIR-MIR-DC12} and \ref{NIR-DC12SFR}),
which detected as a diffuse source in the {\it WISE} image, probably because of the low stellar density of the cluster \citep{Yasui08},
which is similar to that for the Taurus star-forming association.
Figure \ref{MIR-WB89-789} shows the {\it WISE} image of another very distant star-forming region, WB89-789 \citep{Brand07}.
It was also detected as compact reddened sources in the {\it WISE} image.
We did not plot CO distribution superposed on the {\it WISE} image of the WB89-789 cloud, because we do not have CO data
\citep[the distribution of the star-forming region and associated CO clouds are shown in Figures 1, 3, and 6 in][]{Brand07}.

Next, we performed a similar analysis on the molecular clouds with an associated sample star-forming region in the FOG identified by \citet{Snell02}.
Figure \ref{MIR-SnellSFRs} shows the image of CO clouds (left panels) and with {\it WISE} (right panels).
Just like the case of the EOG (Digel Clouds~1 and 2), we found that all of the sample star-forming regions in the FOG were clearly detected as reddened sources
but were not resolved into individual stars with {\it WISE}.
All of them but one were detected in the {\it WISE} images as compact sources.
The sole exception was 2395+6244, which looked like a diffraction spike in the {\it WISE} image.

\subsubsection{WISE sources} \label{sec:3.2.2}
For all of the sample star-forming regions, we searched for corresponding {\it WISE} sources in the AllWISE source catalog (see Section \ref{sec:2.1})
and selected sources within 3$^{\prime \prime}$ from each star-forming region (e.g. yellow circles in Figure \ref{NIR-DC12SFR}).
Among these, the {\it WISE} sources of star-forming regions in 02395+6244 cloud and Cloud~2 were found to have {\it cc\_flag} (see Section \ref{sec:2.1}).
The star-forming regions in  (and other sources around) the 02395+6244 cloud have the D flag due to a very bright {\it WISE} source in the CO peak (blue star symbol in Figure \ref{MIR-SnellSFRs}).
Some obvious star-forming regions in the northern half of Cloud~2 (and other {\it WISE} sources around those star-forming regions)
have the P flag for an unknown reason (blue star symbols in Figure \ref{CO-NIR-MIR-DC12}).
All of the flagged {\it WISE} sources except for the sources in the northern half of Cloud 2 were rejected.
The justification for the exception in the case of Cloud~2 is that we did not recognize any latent features in the {\it WISE} image
and suspect that the AllWISE catalog incorrectly identified the extended photodissociation regions in the northern half of Cloud~2  \citep{Kobayashi08} as a latent feature.
Source-by-source checking of the flagged sources would be beneficial, though labor-intensive, as a future project to obtain the more reliable identifications.
Note that \citet{Koenig14} also checked the reduced chi-square ($\chi^2$) of the profile-fit photometry of the {\it WISE} sources, in addition to the {\it cc\_flags},
and concluded that the photometry of many sources with low S/N and high $\chi^2$, which means extended, is inaccurate \citep[e.g. Figure 1 in][]{Koenig14}.
However, it is impractical in this study to reject all the sources with a high $\chi^2$,  because virtually all of the distant star-forming regions were
identified as extended sources (high $\chi^2$) in the {\it WISE} images.
All the identified {\it WISE} sources are listed in Table \ref{tbl:sampleSFR}.

\subsection{WISE magnitude and color of sample star-forming regions} \label{sec:3.3}
We set a 1$^\circ$ $\times$ 1$^\circ$ area around each of the molecular clouds with associated sample star-forming regions (hereafter the sample area)
and classified all the {\it WISE} sources in the area as star-forming regions or the others, which are the background and foreground objects.
In Figure \ref{Sample-REG}, the red dots show the sources of sample star-forming regions,
and the dots in the other colors show all the other sources in the field.
In this subsection, we investigate the colors and magnitudes of all of the {\it WISE} sources in the area
to construct an identification criterion for unresolved distant star-forming regions.

\subsubsection{Color-magnitude diagram} \label{sec:3.3.1}
Figure \ref{Sample-CM} shows the [3.4] versus [3.4]$-$[4.6] color-magnitude diagram of all of the {\it WISE} sources in the sample area.
The yellow and red symbols show the sample star-forming regions in the FOG and the EOG, respectively.
The yellow star symbols show the star-forming regions known to accompany OB stars in the FOG (see Section \ref{sec:3.1} and Table \ref{tbl:sampleSFR}).
The black dots show all the other sources in the area.
While most other sources are distributed at around [3.4]$-$[4.6] = 0, all of the star-forming regions are at [3.4] $-$ [4.6] $\ge$ 0.5.
The latter color region corresponds to $A_K$ $>$ 2 mag \citep{McClure09,Koenig14},
some of which can be foreground extinction due to the large distance of the star-forming regions,
as well as to the intracluster extinction, given that they are embedded clusters.
In particular, five star-forming regions show much redder colors of [3.4]$-$[4.6] $>$ 1.5  (equivalent extinction of $A_K$ $>$ 7 mag),
which may originate from infrared excess of circumstellar disks or envelopes in addition to the extinction.

We also found in Figure \ref{Sample-CM} that the distributions of [3.4] for star-forming regions in the FOG and EOG are separated with a little overlap:
[3.4] = 9--12 and 11--16 mag ranges, respectively.
If star-forming regions in the FOG and the EOG have a similar intrinsic luminosity, those in the EOG are expected to be about 2 mag fainter,
because their distances are $D$ = 6.5--10 kpc and $D$ = 12--16 kpc, respectively.
The abovementioned distributions are consistent with this expectation.
Note that some of the EOG star-forming regions are even fainter by a further 2 mag,
presumably because most of them are faint-end star-forming regions (stellar aggregates),
which were found only with very deep NIR imaging of Clouds~1 and 2 (see Table \ref{tbl:sampleSFR}).
The limiting magnitude of 3.4 {\micron} is fainter than 16.5 mag \citep{Wright10}.
Therefore, {\it WISE} is confirmed to have enough sensitivity to detect all kinds of star-forming regions up to the edge of the Galaxy.

\subsubsection{Color-color diagram} \label{sec:3.3.2}
Figure \ref{Sample-CC} shows the [3.4]$-$[4.6] versus [4.6]$-$[12] color-color diagram of all of the {\it WISE} sources in the sample area.
The notation is the same as in the [3.4] versus [3.4]$-$[4.6] color-magnitude diagram (Section \ref{sec:3.3.1} and Figure \ref{Sample-CM}).
We also show (black dashed lines) the region of individual YSOs in the solar neighborhood taken from \citet[][e.g., their Figure 5 and Section 1 in this paper]{Koenig14}.
We compared the {\it WISE} colors of the star-forming regions with those of individual YSOs and confirmed that the star-forming regions are primarily distributed in the YSO region
defined by \citet{Koenig14} on the diagram.
In particular, many of the star-forming regions were found to be distributed near and within the Class I YSO region in \citet{Koenig14}.
It is an expected result, because our targets are basically accompanied by and embedded in their parental clouds, and virtually all of them are affected by significant infrared excess from circumstellar disks
and envelopes of Class I sources.
However, several star-forming regions were found to be located outside the YSO regions, in the lower right in the diagram (Figure \ref{Sample-CC}).
It is probably due to a significant amount of polycyclic aromatic hydrogen (PAH) emission, which is known to be strong in the 12 {\micron} and,
to a lesser degree, 3.4 {\micron} bands \citep{Wright10}.
PAH is generally present in star-forming regions with OB stars, whose UV flux induces the PAH emission.
In fact, all of the star-forming regions that are known to have OB stars in our sample were found to be in this PAH excess region,
as marked with yellow star symbols in Figure \ref{Sample-CC}.

\subsubsection{Identification criterion for distant star-forming regions} \label{sec:3.3.3}
Based on the distribution of plots for sample star-forming regions in the [3.4]$-$[4.6] versus [4.6]$-$[12] color-color diagram,
we empirically defined the region for star-forming regions on the diagram:
[3.4]$-$[4.6] $\ge$ 0.5, [4.6]$-$[12] $\ge$ 2.0, and [4.6]$-$[12] $\le$ 6.0 (magenta lines in Figure \ref{Sample-CC}).
We defined this region so that it excludes a part of the region, [3.4]$-$[4.6] = 0.25--0.5 and [4.6]$-$[12] = 1.0--3.0,
in the Class II YSO region by \citet[][black dashed lines in Figure \ref{Sample-CC}]{Koenig14}
to avoid contamination by diskless photospheres (Class III YSOs and unrelated field stars) located in the same region
\citep[left panel of Figure 10 in the][]{Koenig14}.
Since our samples are found to be concentrated near and within the Class I YSO region by \citet[][Section \ref{sec:3.3.2}]{Koenig14},
the drawbacks of eliminating this region are likely to be limited.
In contrast, this defined region encompasses an additional region of [3.4]$-$[4.6] $\ge$ 0.5 and [4.6]$-$[12] = 3.0--6.0,
which is outside the YSO region by \citet{Koenig14}, so that we can pick up the star-forming regions accompanying OB stars with possible PAH emission
at the 3.4 and 12 {\micron} bands, as discussed in Section \ref{sec:3.3.2}.
We cut off at [4.6]$-$[12] = 6.0 to reduce the contamination from planetary nebulae (PNe), which are known to have large values of
[3.4]$-$[4.6] ([3.4]$-$[4.6] = 4--7; left panel of Figure 10 in \citet{Koenig14}.
Furthermore, we set the lower side of the additional region at [3.4]$-$[4.6] = 0.5, considering the potential contamination by starburst galaxies
\citep[e.g. Figure 10 in][and Figure 17 on the website of  the {\it WISE} All-sky Data Products\footnote{\url{http://wise2.ipac.caltech.edu/docs/release/allsky/expsup/sec2_2.html}}]{Wright10}.

\subsection{New candidates for star-forming regions in the sample clouds} \label{sec:3.4}
Using the identification criterion, we searched for new candidate star-forming regions in the molecular clouds with associated sample star-forming regions
where no NIR images are available.
We identified 58 new candidate star-forming regions within the 3 $\sigma$ contours of the molecular clouds (Table \ref{tbl:sampleSFR}, yellow star symbols in Figures
\ref{CO-NIR-MIR-DC12} and \ref{MIR-SnellSFRs}, and yellow dots in Figure \ref{Sample-REG}).

\subsection{Possible contamination} \label{sec:3.5}
Here we discuss the possible contamination by foreground and/or background objects in the region defined in Section \ref{sec:3.3.3} on the color-color diagram for star-forming regions.
The defined region for star-forming regions is contaminated mainly by YSOs associated with foreground molecular clouds, foreground/background PNe,
and background active galactic nuclei (AGNs) and quasi-stellar objects (QSOs) \citep[e.g.][]{Koenig14,Wright10}.
Indeed, we found clear contamination by possible AGNs near the CO peaks of Cloud~2 (see Figures \ref{CO-NIR-MIR-DC12} and \ref{DC2_Con});
although the {\it WISE} sources showed the color that is typical for star-forming regions, the corresponding objects are recognized as three galaxies,
all of which have a symmetric disk and bright pointlike core, in the high-resolution Subaru NIR images (FWHM $\sim$ 0.$^{\prime \prime}$3--0.$^{\prime \prime}$35),
and hence are very likely to be AGNs.

To estimate the total contamination rate quantitatively, we compared the number density of the candidate (and sample) star-forming regions in the cloud region
(inside the 3 $\sigma$ contour of molecular clouds; $n_{\rm MC}$)
with those in the field region (the circular region within the radius 0$^\circ$.5, center at the CO peak, with the cloud region excluded; $n_{\rm F}$).
We calculated $n_{\rm MC}$ and $n_{\rm F}$, using the numbers of candidate (and sample) star-forming regions in the cloud and the field regions ($N_{\rm MC}$ and $N_{\rm F}$, respectively)
and the areas of the cloud region ($A_{\rm MC}$), which were manually estimated on the image, and of the field region ($A_{\rm F}$ = 0.5$\times$0.5$\times$$\pi$ $-$ $A_{\rm MC}$  deg$^2$).
Then, we calculated contamination rates of 100 $\times$ $n_{\rm F}/n_{\rm MC}$.
Table \ref{tbl:sampleContami} lists the contamination rates for all the clouds with associated sample star-forming regions
except the WB89-789 cloud\footnote{We did not estimate the contamination rate for the WB89-789 cloud because we do not have CO molecular data for it.}.
We found the median number density of candidates in the field region and contamination rate to be $\sim$ 1.5 $\times$ 10$^{-2}$ arcmin$^{-2}$ and $\sim$ 10 \%, respectively,
the latter of which is equivalent to roughly one contamination source in each molecular cloud with associated sample star-forming regions.

We suspect that the main contamination source is likely to be YSOs associated with foreground molecular clouds, because the number densities of the other possible contamination sources
(PNe, AGNs, and QSOs) are lower by an order of magnitude of 1 or more than the median number densities of candidates in the field region.
The number densities of PNe and that of AGNs and QSOs combined are estimated to be only 3.0 $\times$ 10$^{-4}$ and 4.2 $\times$ 10$^{-3}$ arcmin$^{-2}$, respectively,
based on the expected Galactic PN population of 30,000--50,000 \citep[e.g.][]{Frew06,Frew08,Moe06} for the former
and the expected number of AGNs and QSOs detected with {\it WISE} \citep[Figure 16 in][]{Wright10} for the latter.
In fact, the contamination rate appeared to depend highly on the amount of foreground molecular clouds around the target molecular clouds with associated star-forming regions.
As an example, Figure \ref{Foreground-clouds} shows the distribution of foreground molecular clouds in and around representative molecular clouds with associating star-forming regions:
Cloud~2 with a contamination rate of $\sim$3 \% (left panel) and the 02376+6030, 02407+6029, and 02413+6037 clouds with contamination rate of 20--60 \% (right panel).
In Figure~\ref{Foreground-clouds}, green contours show the CO distribution of Cloud~2 and the 02376+6030, 02407+6029, and 02413+6037 clouds,
and the other color contours show the CO distribution of foreground molecular clouds.
Figure \ref{Foreground-clouds} demonstrates that the amount of foreground molecular clouds in and around molecular clouds with associated star-forming regions
with a large contamination rate appears to be larger than that with a low contamination rate.
Furthermore, the contamination rates also depend on the spatial resolution of the CO map.
For example, the contamination rate of Cloud~1 decreases from 6 \% to 4 \%
if the CO data with the FCRAO CO survey of the outer Galaxy \citep[resolution $\sim$ 100$^{\prime \prime}$;][]{Brunt03}
are replaced with those with our NRO 45 m radio telescope \citep[resolution $\sim$ 17$^{\prime \prime}$;][]{Izumi14},
as the total area of the identified cloud decreases.
It would be useful to keep compiling high-resolution CO data of those distant molecular clouds for better identification.

\section{A new survey of star-forming regions in the outer Galaxy} \label{sec:4}
\subsection{New candidate of star-forming regions} \label{sec:4.1}
Applying the methods described in Section \ref{sec:3}, here we systematically search the {\it WISE} data in 466 BKP clouds in the outer Galaxy (see Section \ref{sec:2.2})
for new star-forming regions.
We selected AllWISE catalog sources that do not have {\it cc\_flags} in all of the four bands and have S/N $>$ 5 in all of the 3.4, 4.6, and 12 {\micron} bands (see Section \ref{sec:2.1})
in the FCRAO CO survey of the outer Galaxy (926,132 sources in the survey area: 102$^\circ$.49 $\le$ $l$ $\le$ 141$^\circ$.54 and $-$3$^\circ$.03 $\le$ $b$ $\le$ 5$^\circ$.41).
From the [3.4]$-$[4.6] versus [4.6]$-$[12] color-color diagram of the sources (Figure \ref{CC_FCRAO}),
we found 778 candidate star-forming regions within 3 $\sigma$ contours of 252 clouds with a galactocentric radius of up to $R_{\rm G}$ $\sim$ 20 kpc
\footnote{We found 11 candidates with kinematic distance of more than 20 kpc ($R_{\rm G}$  $>$ 27 kpc), but they have already been found (star forming regions associated with Digel Cloud 2) and are
known to be actually located at a distance of 12 kpc ($R_{\rm G}$  = 19 kpc) from high-resolution optical spectra \citep[e.g.][]{Smartt96, Kobayashi08}.}.
Among these candidates, 67 candidates in 12 clouds were already found in the previous section (Table \ref{tbl:sampleSFR}, Figures \ref{CO-NIR-MIR-DC12} and \ref{MIR-SnellSFRs}).
Note that more than one cloud at different velocities exists in the line of sight for 13 out of the 778 candidates, and we cannot determine which cloud is parental for them.
All of the candidates and their parental clouds are listed in the Appendix.

\subsection{Contamination rate} \label{sec:4.2}
Here we examine how likely it is that the selected candidates are genuine star-forming regions.
We estimated the contamination rates for the 252 clouds that have associated candidate star-forming regions using the same method as in Section \ref{sec:3.5}.
In this section, for the sake of simplicity, we set a field region around the clouds as an annulus instead of the field region used in Section \ref{sec:3.5}
(the circular region with a radius of 0$^\circ$.5 centered at the CO peak, with the cloud region excluded).
We set the inner radius of the annulus as twice the radius estimated from the cloud region,
2 $\times$ $\sqrt{A_{\rm cloud}/\pi}$, where $A_{\rm cloud}$ is derived from the number of spatial pixels from the BKP catalog with the pixel scale of 0$^{\prime}$.837.
The mean derived inner radius of the annulus was 0$^\circ$.24.
Note that whereas 4.7$\sigma$ is the threshold used both for the identification of the original BKP clouds and for the boundary for the area $A_{\rm cloud}$,
we adopted the lower threshold contours of 3$\sigma$ in searching for candidate star-forming regions in order to pick up as many star-forming regions as possible,
considering that they were sometimes located near the edge of a cloud (in the vicinity of 3 $\sigma$ contours; e.g. Figure \ref{CO-NIR-MIR-DC12}).
Thus, the absolute value of the contamination rate for the cloud region within the 3$\sigma$ contour must be slightly higher than that within the 4.7$\sigma$ contour,
but the difference should be insignificant, given the small area difference between 3$\sigma$ and 4.7$\sigma$
(e.g. see Figure \ref{MIR-SnellSFRs} for the distribution of 3$\sigma$ and 5$\sigma$ CO contours in several regions).

To set the appropriate field region, we investigated the variation in number density of candidates in trial field regions (top panel of Figure \ref{ND-CR})
and of the contamination rate (bottom panel of Figure \ref{ND-CR}) for widths of the annulus varied from 0$^\circ$.05 to 1$^\circ$.0.
We found in Figure \ref{ND-CR} that both the number densities of candidate star-forming regions in the field region and contamination rate of each cloud were
roughly constant for widths of annulus between 0$^\circ$.4 and 1$^\circ$.0
and start to vary at a width of roughly 0$^\circ$.3 or lower, probably because the area with that width is so small that
the derived parameters were affected by a local distribution of candidate star-forming regions.
Thus, we set the width of the annulus to be 0$^\circ$.5 to avoid the effect of local distribution of candidate star-forming regions.

Between the mean number density of candidate star-forming regions in the field region (0.020 arcmin$^{-2}$; cyan dashed line in Figure \ref{ND-CR})
and total number density of QSO, AGNs, and PNe (0.0045 arcmin$^{-2}$; green dashed line),
the former is found to be about five times higher than the latter.
Therefore, the main contamination source is likely to be YSOs associated with foreground molecular clouds, and it is consistent with our previous result with the sample region (Section \ref{sec:3.5}).
As expected, the mean number density of candidate star-forming regions in the cloud regions (0.13 arcmin$^{-2}$ for each; red dashed line in Figure \ref{ND-CR})
was much higher than that in the field regions (cyan dashed line).

Figure \ref{Contami} shows the distribution of contamination rates with the width of the annulus of 0$^\circ$.5.
We found that about 20 molecular clouds have a high contamination rate ($\ge$ 50\%), whereas almost all of the clouds ($\sim$ 230 molecular clouds)
have a contamination rate of lower than 50\%.
We also found that the number of molecular clouds sharply drops at a contamination rate of 20--30\% (Figure \ref{Contami}).
Therefore, we set the upper contamination threshold to be 30\% for a candidate to be regarded as reliable.
Then, 211 out of 252 molecular clouds with associated candidate star-forming regions satisfy this condition.
This number is large enough for statistical studies of star-formation activity in the outer Galaxy.

\subsection{Properties of candidate star-forming regions} \label{sec:4.3}
To investigate properties of the candidate star-forming regions, we first checked the distribution of the candidates in the
[3.4]$-$[4.6] vs. [4.6]$-$[12] color-color diagram (Figure \ref{CC_FCRAO})
and [3.4] versus [3.4]$-$[4.6] color-magnitude diagram (Figure \ref{CM_FCRAO}).
In those figures, the red circles and yellow squares show the candidate star-forming regions with a contamination rate of their parental clouds of $<$ 30\% and $\ge$ 30\%, respectively.
The black dots in the left panels and black contours in the right panels show the distribution of all 926,132 sources in the area.
We found that 26\% (204/778) of the candidates are located outside the YSO region by \citet{Koenig14},
as indicated with cyan dashed lines in Figure \ref{CC_FCRAO}.
This suggests that those ``outsiders'' could be relatively massive star-forming regions (including OB stars; see Section \ref{sec:3.3.2}).
We found no significance difference in the distributions between the candidates with a contamination rate of their parental clouds of $<$ 30\% and $\ge$ 30\% in either of Figures \ref{CC_FCRAO} and \ref{CM_FCRAO}.

Next, we checked the relation between the magnitude of the candidates in all the four bands and the kinematic distances of their parental clouds  (Figure \ref{Mag-D}).
The left panel of Figure \ref{Mag-D} shows the variation of apparent magnitudes as a function of the kinematic distance.
The notation is the same as in Figures \ref{CC_FCRAO} and \ref{CM_FCRAO}.
The gray dashed lines show the average detection limit for the minimum integration for eight frames
\citep[16.5, 15.5, 11.2, and 7.9 mag for 3.4, 4.6, 12, and 22 {\micron}, respectively;][]{Wright10}.
The apparent magnitudes spread widely by 4 to 5 orders.
The apparent magnitude difference between the brightest candidates with the distances of $D$ = 6 and 15 kpc was found to be about 2 mag, which is consistent with the one calculated from the distance
$\Delta m$ = 5 $\times$ $\log_{10}$(15/6) = 2, assuming that the absolute magnitude of the brightest star-forming regions is constant.
The candidates with a contamination rate of their parental cloud of higher than 30 \% (yellow squares in Figure \ref{Mag-D})
are concentrated at 6 kpc $\le$ $D$ $\le$ 10 kpc.
This suggests that the major factor that affects the contamination rate is the apparent size of the cloud,
which is larger for clouds at a smaller distance.

In the left panel of Figure \ref{Mag-D}, we also show the apparent magnitudes of A0 and B0 stars in the main sequence for all the four bands \citep[cyan and blue curves; calculated based on Tabled 7.5 and 15.7 in][]{Cox00}.
Note that we used intrinsic $V$ $-$ $L$ colors ($V$ band: 0.5555 {\micron}; $L$ band: 3.547 {\micron}) and the absolute $V$-band magnitudes of A0 and B0 stars for calculating
the apparent magnitudes of those stars for any of the four bands,
since infrared colors, such as [3.4]$-$[22] are negligible for those early-type stars.
We also superposed in Figure \ref{Mag-D} the 22 {\micron} magnitude for  \ion{H}{2} regions ionized by B0 stars (blue dotted curve)
derived from Figures 1 and 2 in \citet{Anderson14} with a zero-magnitude flux density of $F_{\nu\, \rm 22 \micron}$ = 8.363 Jy \citep{Jarrett11}.
At 3.4 and 4.6 {\micron}, most of the apparent magnitudes of the {\it WISE} sources are within the range of the typical values for
stellar aggregates (single A-type star $+$ T Tauri stars) to OB associations,
whereas those at 12 and 22 {\micron} show much lower magnitudes.
This result suggests that magnitudes of 3.4 and 4.6 {\micron} are dominated by stars,
whereas magnitudes of 12 and 22 {\micron} are dominated by dust emission from circumstellar dust and/or free-free emission ($+$ PAH emission) from \ion{H}{2} regions.
This is consistent with the well-known SED of star-forming regions (see Section \ref{sec:2.1}).
Consequently, we confirm that we can detect stellar aggregates up to $D$ $\sim$ 10 kpc ($R_{\rm G}$ $=$ 15--17 kpc) with the {\it WISE} data.

The right panel of Figure \ref{Mag-D} shows the variation of the absolute magnitudes calculated from the {\it WISE} magnitudes in the AllWISE Source Catalog
as a function of the kinematic distance of their parental clouds.
The notation is the same as in the left panel of Figure \ref{Mag-D}.
The larger the distance, the higher the detection limit is in the absolute magnitude (gray dashed lines in Figure \ref{Mag-D}).
At the most distant region  ($D$ $\sim$ 14 kpc; $R_{\rm G}$ $\sim$ 20 kpc),
the data set of the {\it WISE} sources was found to be complete for the sources brighter than about
1.0, 0.0, $-$4.0, and $-$7.0 mag at 3.4, 4.6, 12, and 22 $\micron$, respectively (gray solid lines in right panels of Figure \ref{Mag-D}).

\subsection{Comparison with {\it IRAS} point sources} \label{sec:4.4}
\citet{Kerton03} searched the {\it IRAS} database for the point sources associated with BKP clouds and found 96 {\it IRAS} sources in 89 BKP clouds at $R_{\rm G}$ $\ge$ 13.5 kpc.
In order to take into consideration the contamination by foreground or background sources, they used the expected number of associations $N_E$, 
which will associate with a randomly chosen position within the FCRAO CO survey of the outer Galaxy boundaries \citep{Brunt03}. 
They made a color-color diagram of those {\it IRAS} sources and found that the {\it IRAS} sources with a low $N_E$ are located in and around
the color region of the representative star-forming regions by \citet{Wouterloot89}:
$\log{(25 \micron/12 \micron)}$ $>$ 0, 0.38 $<$ $\log{(60 \micron/25 \micron)}$ $<$ 1.88, and $-$0.77 $<$ $\log{(100 \micron/60 \micron)}$ $<$ 0.39 + 0.23 $\times$ $\log{(60 \micron/25 \micron)}$
\citep[e.g. Figure 5 in][]{Kerton03}.
Their result is plotted in Figure \ref{IRAS-CC}, which shows a $\log{(60 \micron/25\micron)}$ vsersus $\log{(25 \micron/12 \micron)}$ color-color diagram
of 34/96 {\it IRAS} sources with $N_E$ $\le$ 10$^{-2.0}$ at $R_{\rm G}$ $\ge$ 13.5 kpc.
The {\it IRAS} color region by \citet{Wouterloot89} is indicated by the box region in Figure \ref{IRAS-CC}.

In our survey with the {\it WISE} data, 24 out of the 34 {\it IRAS} sources were detected as candidate star-forming regions.
In Figure \ref{IRAS-CC}, the filled and open circles indicate the {\it IRAS} sources detected and not detected, respectively,
as candidate star-forming regions in our survey with the {\it WISE} data.
Thus, we confirmed that our identification criterion, which does not use the data at wavelengths longer than 12 {\micron},
is roughly consistent with the criterion by \citet{Wouterloot89}, which uses up to 100 {\micron} data.

\section{Summary} \label{sec:5}
We developed an identification criterion of unresolved distant star-forming regions with the MIR all-sky survey data by {\it WISE}:
[3.4]$-$[4.6] $\ge$ 0.5, [4.6]$-$[12] $\ge$ 2.0, and [4.6]$-$[12] $\le$ 6.0.
The criterion enables us to pick up star-forming regions effectively by combining it with CO survey data in the outer Galaxy.
We applied the criterion to 466 molecular clouds in the outer Galaxy at $R_{\rm G}$ $\ge$ 13.5 kpc
detected from the FCRAO CO survey of the outer Galaxy (survey region: 120$^\circ$.49 $\le$ $l$ $\le$ 141$^\circ$.54, $-$3$^\circ$.03 $\le$ $l$ $\le$ 5$^\circ$.41)
and identified 788 candidate star-forming regions in 252 clouds at a galactocentric radius of up to $R_{\rm G}$ $\sim$ 20 kpc.
Among the 788 candidates, 711 in 240 clouds were newly identified.
Comparing the flux densities of the candidates with those of A0 and B0 stars in the main sequence,
we confirmed that we can detect stellar aggregates (single A-type star $+$ T Tauri stars) up to $D$ $\sim$ 10 kpc ($R_{\rm G}$ = 15--17 kpc) with the {\it WISE} data.
Our results imply that our criterion with the {\it WISE} data is very powerful for detecting distant star-forming regions.
With this greatly increased number of candidates, one can now perform statistical studies of star-formation properties in the outer Galaxy for the first time.
We will discuss the distribution and properties of these newly identified star-forming regions in our next paper.

\acknowledgments
This publication makes use of data products from the {\it WISE}, which is a joint project of the University of California, 
Los Angeles, and the Jet Propulsion Laboratory (JPL)/California Institute of Technology (Caltech), funded by the National Aeronautics and Space Administration (NASA).
The research presented in this paper has used data from the Canadian Galactic Plane Survey, a Canadian project with international partners, supported by the Natural Sciences and Engineering Research Council.
This research has made use of the NASA/IPAC Infrared Science Archive, which is operated by the JPL, Caltech, under contract with the NASA.
This research made use of Montage, funded by the NASA's Earth Science Technology Office, Computational Technologies Project, under Cooperative Agreement Number NCC5-626 between NASA and the Caltech. The code is maintained by the NASA/IPAC Infrared Science Archive.
This research used the facilities of the Canadian Astronomy Data Centre operated by the National Research Council of Canada with the support of the Canadian Space Agency.
We thank the anonymous reviewer for a careful reading and thoughtful suggestions that significantly improved this paper.
This study is financially supported by KAKENHI (20340042) Grant-in-Aid for Scientific Research (B) and KAKENHI (26800094) Grad-in-Aid for Young Scientists (B). N.K. is supported by JSPS-DST under the Japan-India Science Cooperative Programs during 2013-2015 and 2016-2018.
\begin{figure*}
\epsscale{0.8}
\plotone{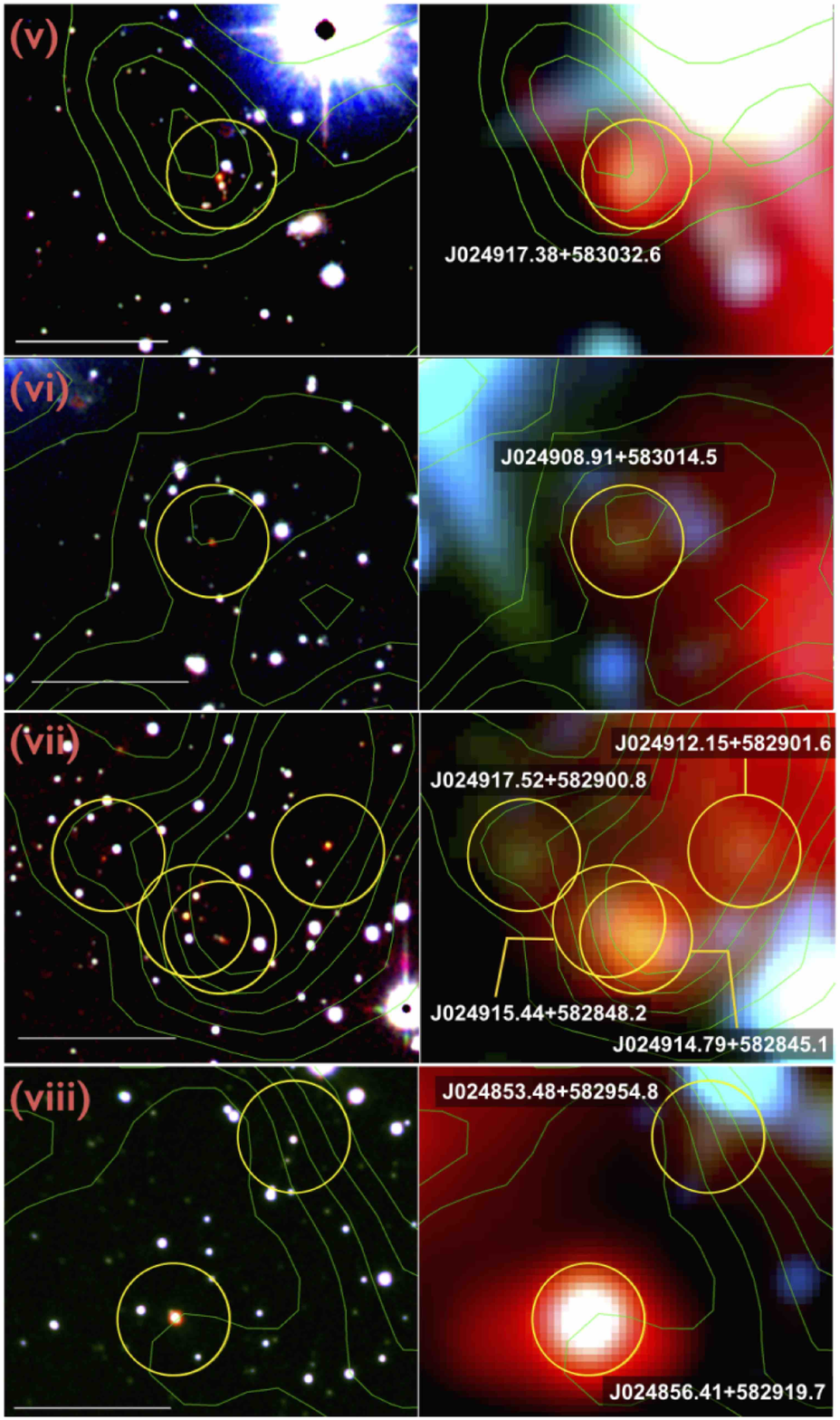}\\
Figure \ref{NIR-DC12SFR}. (Continued.)
\end{figure*} 
\begin{figure*}
\epsscale{0.8}
\plotone{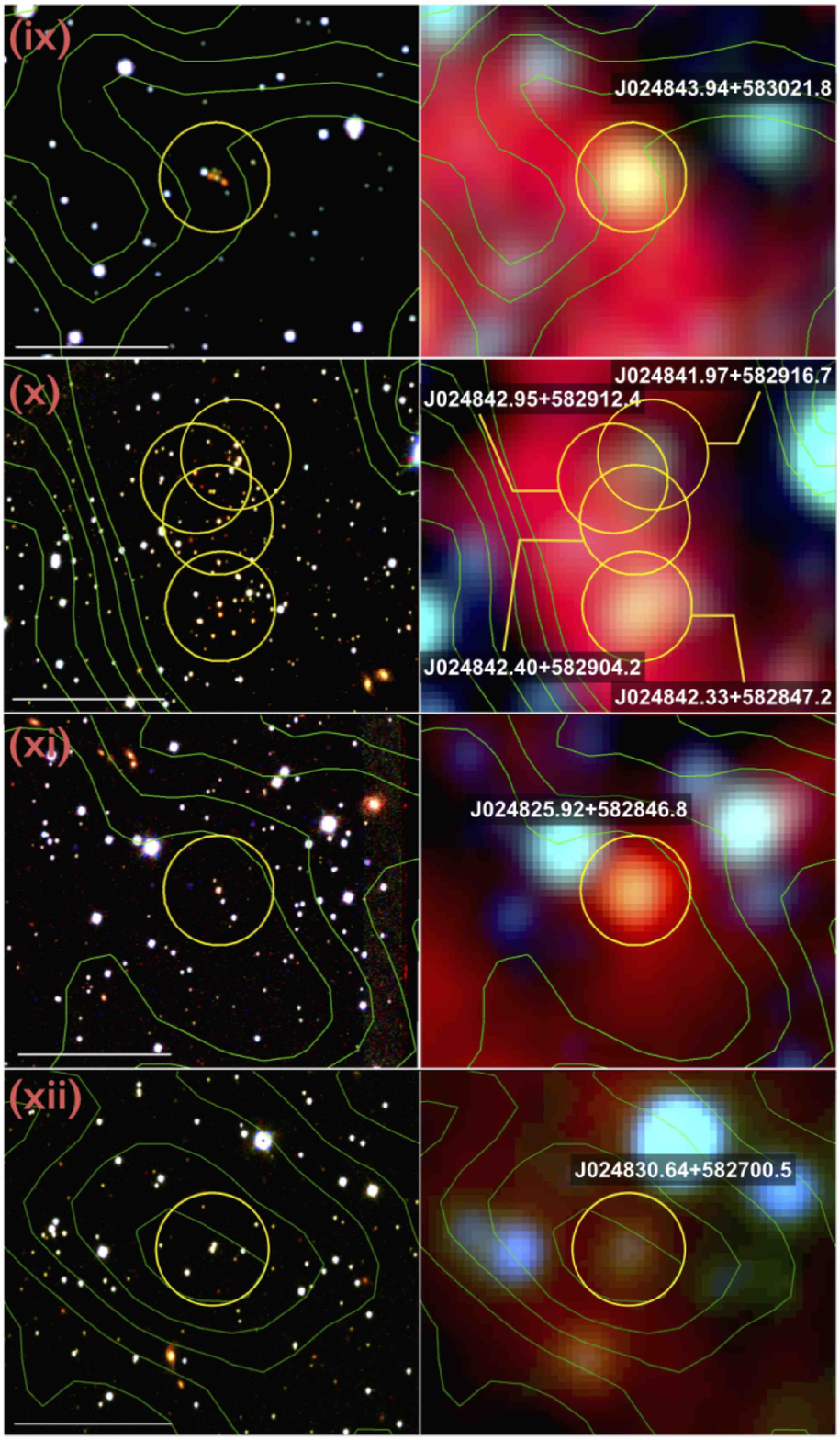}\\
Figure \ref{NIR-DC12SFR}. (Continued.)
\end{figure*} 
\begin{figure*}
\epsscale{0.8}
\plotone{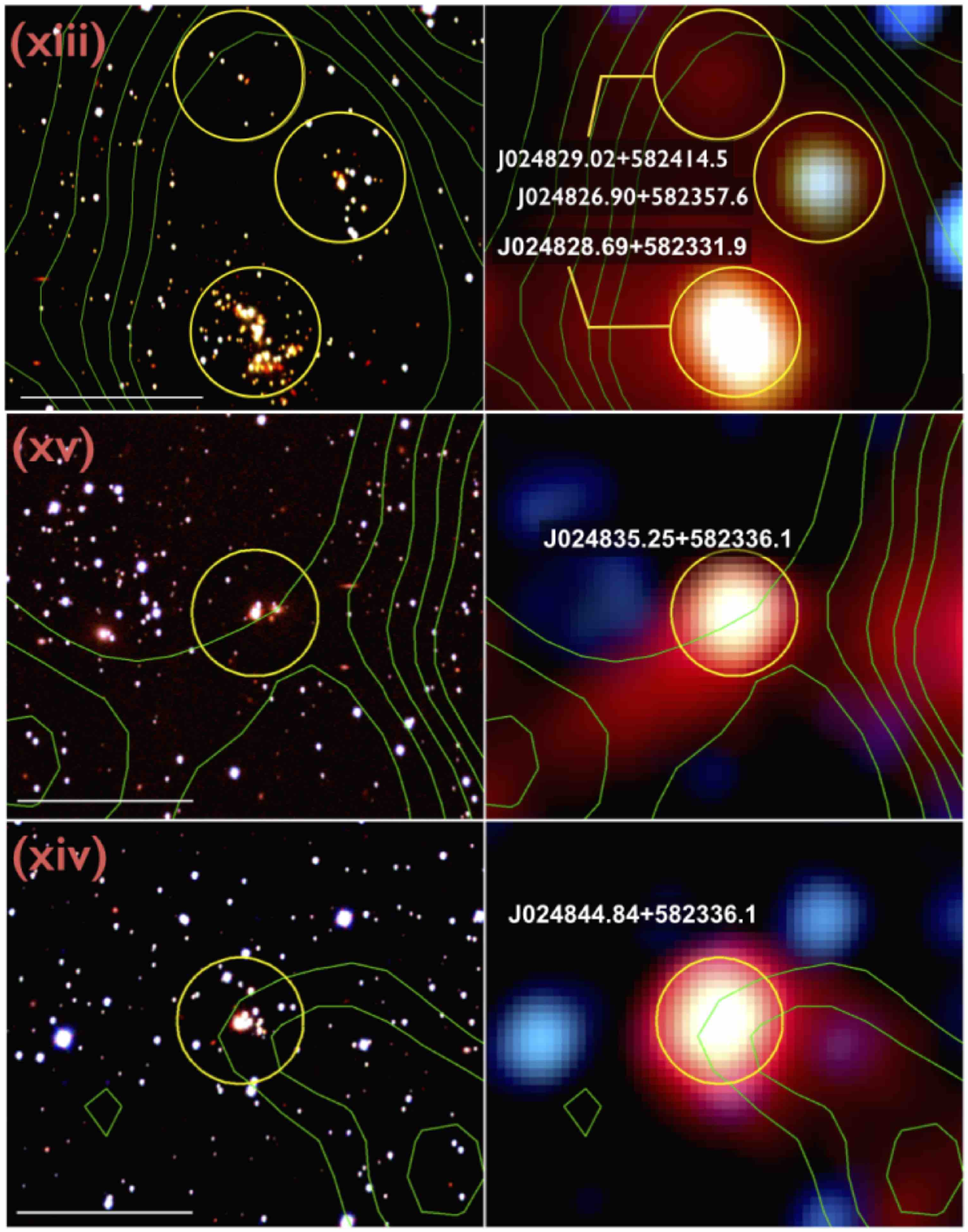}\\
Figure \ref{NIR-DC12SFR}. (Continued.)
\end{figure*} 
\begin{figure*}
\epsscale{0.8}
\plotone{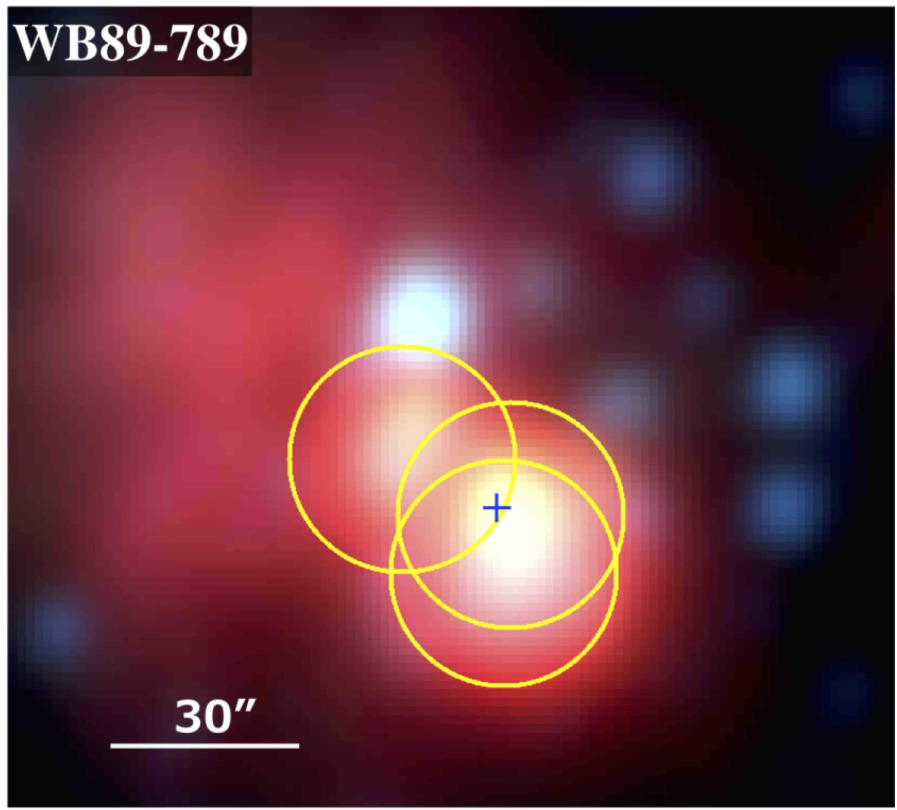}
\caption{{\it WISE} 3.4, 4.6, and 12 {\micron} pseudo-color image of  WB89-789.
Yellow circles show the locations of the {\it WISE} sources for embedded clusters identified by \citet{Brand07}.
The blue cross shows the location of WB89-789 ($l$ = 195$^\circ$.82, $b$ = $-$0$^\circ$.57).}  
\label{MIR-WB89-789} 
\end{figure*}
\begin{figure*}
\epsscale{0.9}
\plotone{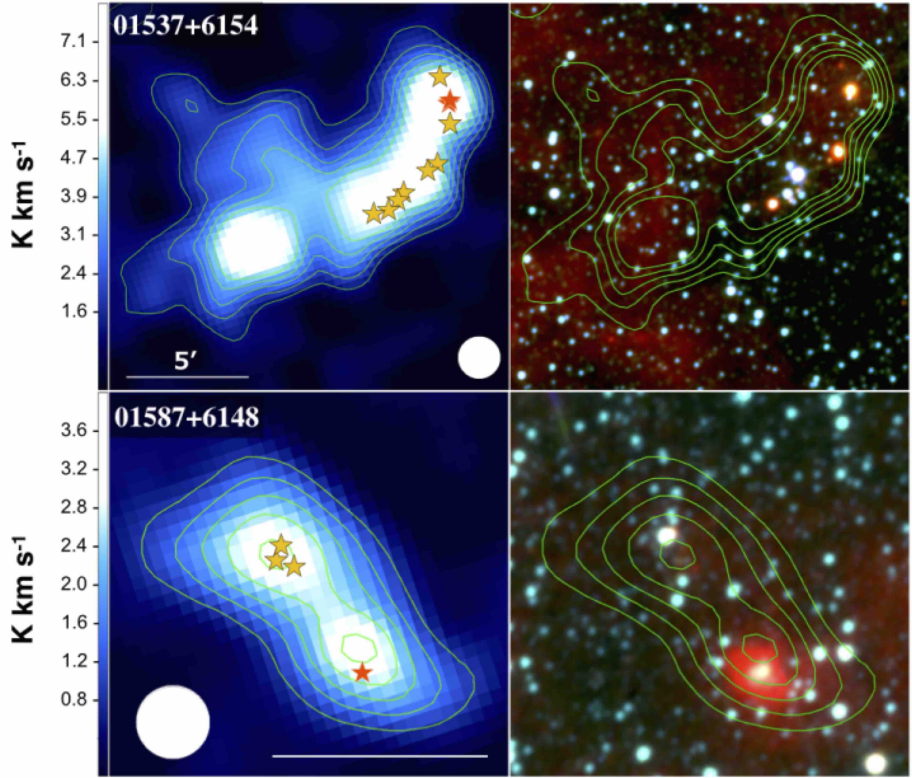}
\caption{
Left: $^{12}$CO(1-0) distribution of the CO peaks in the molecular clouds with associated star-forming regions in the FOG
(based on the CO survey of the outer Galaxy with the FCRAO 14 m telescope, reprocessed by \citet{Brunt03};
01537+6154: $v_{\rm LSR}$ = $-$57.5 to $-$64.9 km s$^{-1}$;
01587+6148: $v_{\rm LSR}$ = $-$59.2 to $-$63.3 km s$^{-1}$; 
02071+6235: $v_{\rm LSR}$ = $-$76.5 to $-$80.6 km s$^{-1}$; 
02376+6030: $v_{\rm LSR}$ = $-$74.0 to $-$79.8 km s$^{-1}$; 
02383+6241: $v_{\rm LSR}$ = $-$69.9 to $-$74.0 km s$^{-1}$; 
02393+6244: $v_{\rm LSR}$ = $-$69.9 to $-$74.0 km s$^{-1}$; 
02407+6029: $v_{\rm LSR}$ = $-$72.4 to $-$78.1 km s$^{-1}$;
02413+6037: $v_{\rm LSR}$ = $-$59.2 to $-$64.1 km s$^{-1}$; 
02421+6233: $v_{\rm LSR}$ = $-$69.9 to $-$74.8 km s$^{-1}$; 
02593+6008: $v_{\rm LSR}$ = $-$57.5 to $-$61.6 km s$^{-1}$).
Green contours show the $^{12}$CO(1-0) distribution with contour levels of 3$\sigma$, 5$\sigma$, 7$\sigma$, 9$\sigma$, and 11$\sigma$
(01537+6154: 1 $\sigma$ = 0.40 K km s$^{-1}$;
01587+6148: 1 $\sigma$ = 0.28 K km s$^{-1}$;
02071+6235: 1 $\sigma$ = 0.28 K km s$^{-1}$; 
02376+6030: 1 $\sigma$ = 0.35 K km s$^{-1}$; 
02383+6241: 1 $\sigma$ = 0.28 K km s$^{-1}$; 
02393+6244: 1 $\sigma$ = 0.28 K km s$^{-1}$; 
02407+6029: 1 $\sigma$ = 0.35 K km s$^{-1}$; 
02413+6037: 1 $\sigma$ = 0.31 K km s$^{-1}$; 
02421+6233: 1 $\sigma$ = 0.31 K km s$^{-1}$; 
02593+6008: 1 $\sigma$ = 0.28 K km s$^{-1}$).
White filled circles show the spatial resolution of the FCRAO data (100.$^{\prime \prime}$88) reprocessed by \citet{Brunt03}.
The red and yellow star symbols show the locations of star-forming regions identified by \citet{Snell02}
and of new candidate ones identified with the {\it WISE} data in this study, respectively.
The blue star symbol shows the location of the known embedded cluster in 02395+6244,
which was not used in this study because of D flags (see Sections \ref{sec:2.1} and \ref{sec:3.2.2}).
Right: {\it WISE} 3.4, 4.6, and 12 {\micron} pseudo-color images of FCRAO star-forming regions in the FOG.
Green contours are identical to those in the corresponding left panel.
}  
\label{MIR-SnellSFRs}
\end{figure*}
\begin{figure*}
\epsscale{0.9}
\plotone{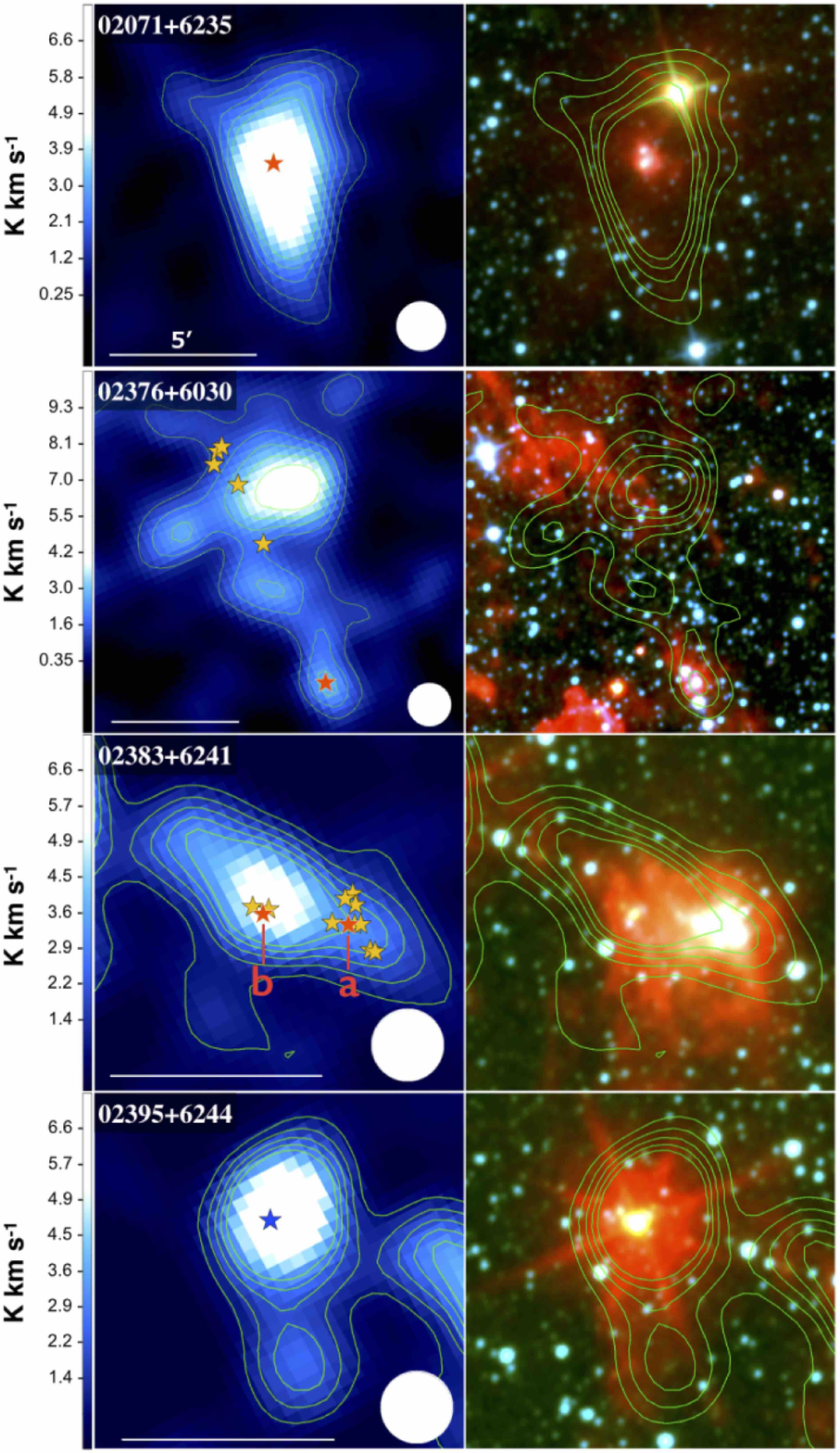}\\
Figure \ref{MIR-SnellSFRs}. (Continued.)
\end{figure*}
\begin{figure*}
\epsscale{0.9}
\plotone{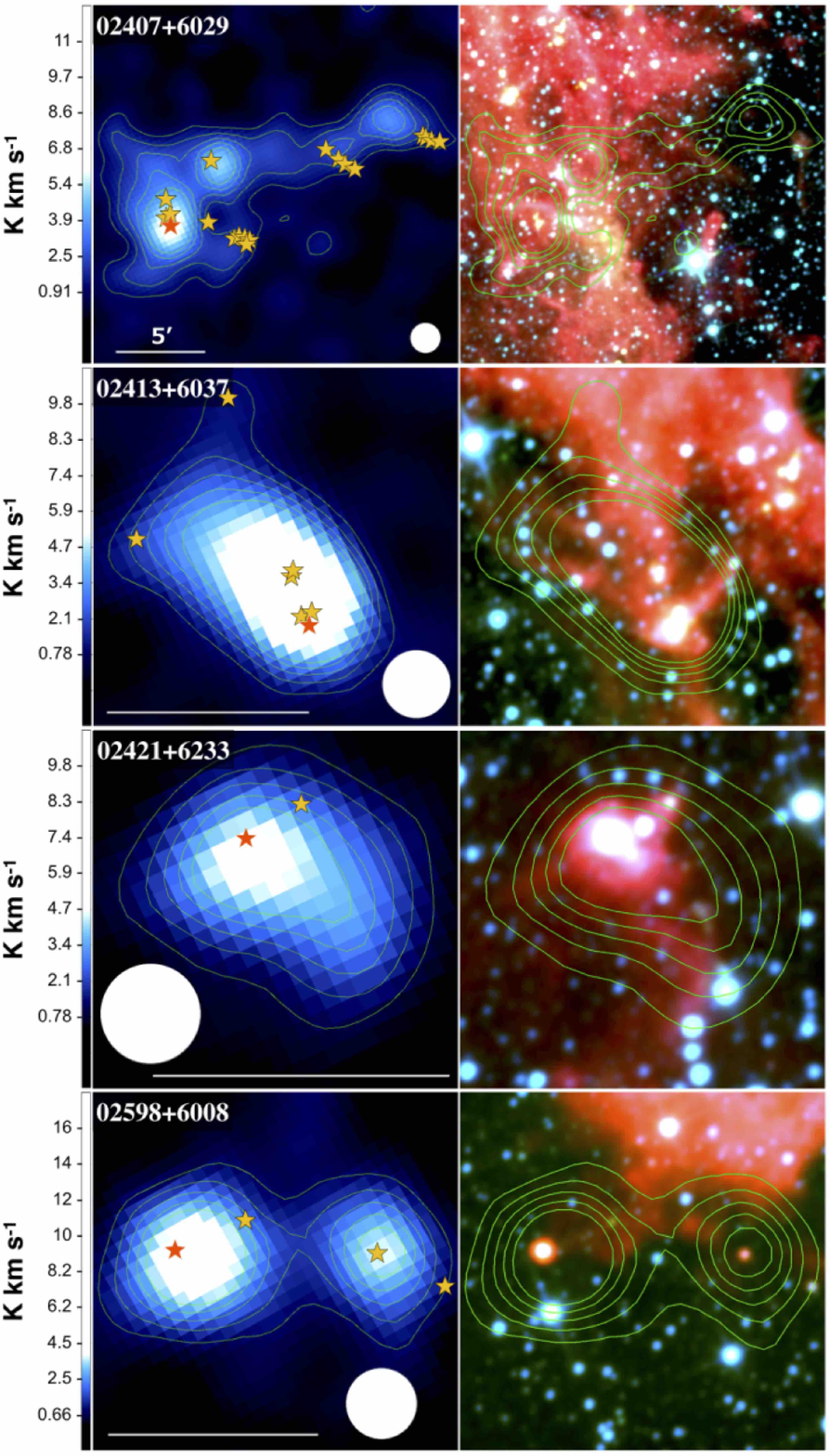}\\
Figure \ref{MIR-SnellSFRs}. (Continued.)
\end{figure*}
\begin{figure*}
\epsscale{1.0}
\plotone{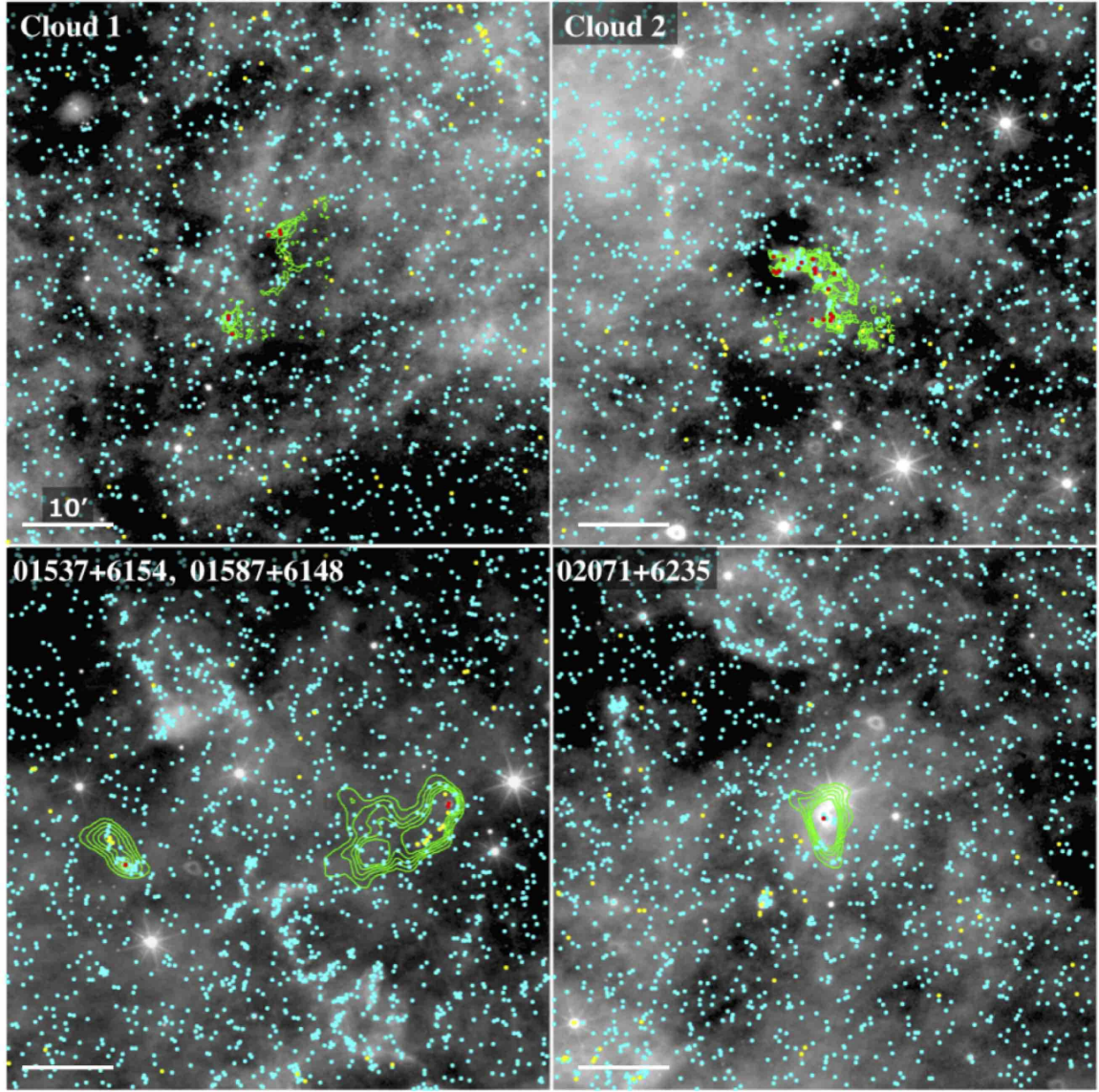}
\caption{
{\it WISE} sources in 1$^\circ$ $\times$ 1$^\circ$ fields in and around the molecular clouds with associated sample star-forming regions (except WB89-789),
plotted on the {\it WISE} 12 $\micron$ gray-scale image.
Red, yellow, and cyan dots show the sample star-forming regions, candidate star-forming regions, and all the other sources in the fields, respectively.
Green contours show the $^{12}$CO(1-0) distribution from the FCRAO data with contour levels of 3$\sigma$, 5$\sigma$, 7$\sigma$, 9$\sigma$, and 11$\sigma$,
the same as in Figure \ref{CO-NIR-MIR-DC12} (Clouds 1 and 2) and Figure \ref{MIR-SnellSFRs} (the other clouds). 
}
\label{Sample-REG} 
\end{figure*}
\begin{figure*}
\epsscale{1.0}
\plotone{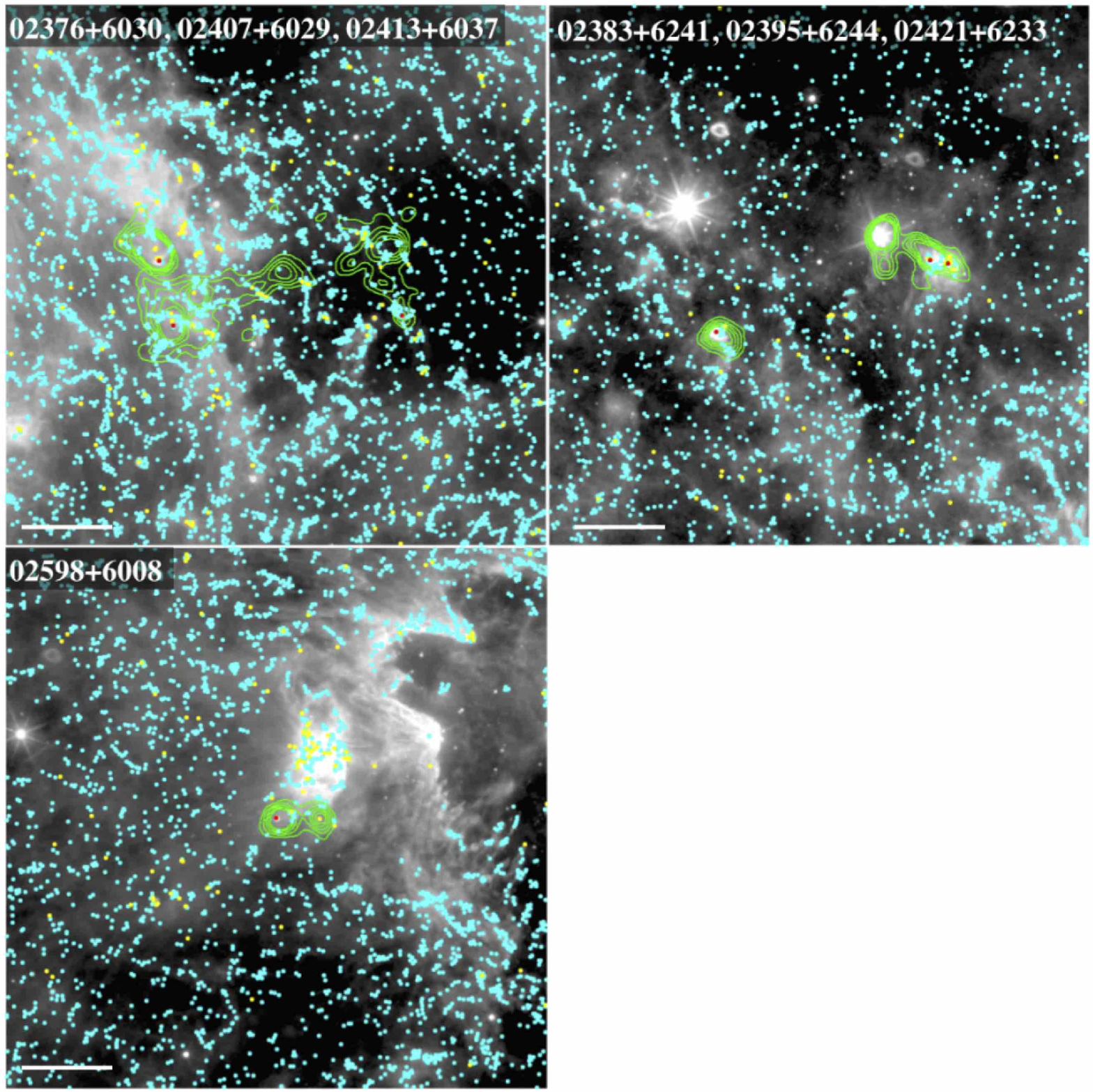}
\\
Figure \ref{Sample-REG}. (Continued.)
\end{figure*}
\begin{figure}
\epsscale{1.2}
\plotone{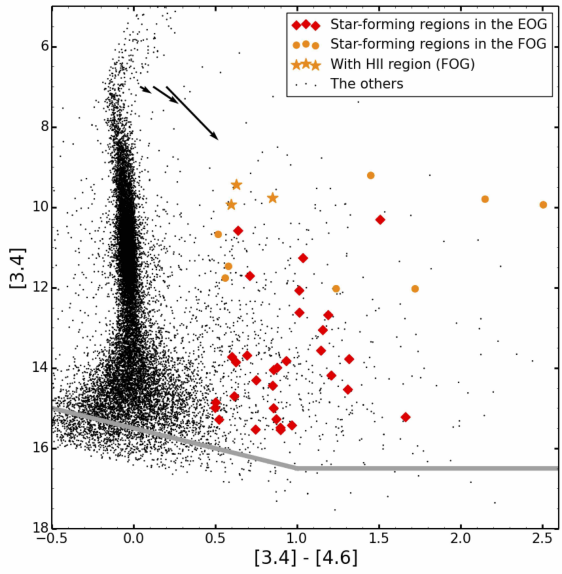}
\caption{
The [3.4] vs. [3.4]$-$[4.6] color-magnitude diagram of the AllWISE catalog source in 1$^\circ$ $\times$ 1$^\circ$ fields in and around molecular clouds with associated sample star-forming regions.
The red and yellow symbols show the sample star-forming regions in the EOG and FOG, respectively.
The yellow star symbols show star-forming regions known to accompany OB stars in the FOG.
Black dots show all the other sources in the fields.
The gray line shows the average detection limit for the minimum integration of eight flames
\citep[16.5 mag for 3.4 {\micron} and 15.5 mag for 4.6 {\micron};][]{Wright10}.
The black arrows shows the extinction vectors
for, from left to right, $A_{K}$ = 0.4, 0.8, and 2 \citep{Koenig14,McClure09}.
} 
\label{Sample-CM}
\end{figure}
\begin{figure}
\epsscale{1.2}
\plotone{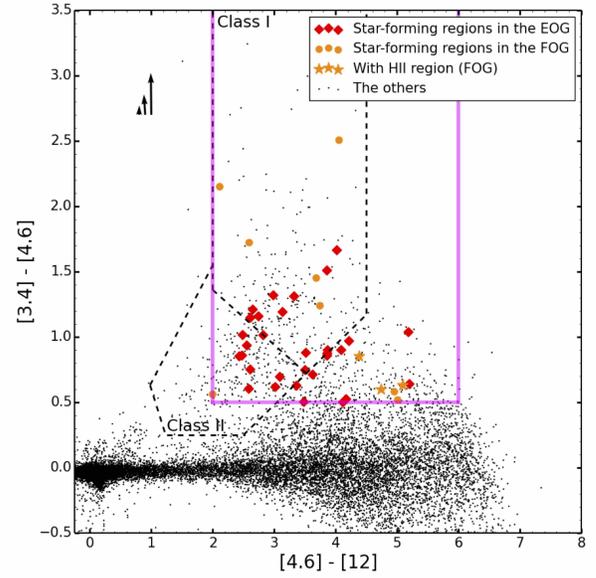}
\caption{
The [3.4]$-$[4.6] vs. [4.6]$-$[12] color-color diagram of the AllWISE catalog sources in 1$^\circ$ $\times$ 1$^\circ$ fields in and around molecular clouds with associated sample star-forming regions.
The red and yellow symbols, yellow star symbols, black dots, and the black arrows are the same as those in Figure \ref{Sample-CM}.
The black dashed lines show the region for typical YSOs \citep{Koenig14}.
The magenta line box shows our defined region for distant star-forming regions.}
\label{Sample-CC}
\end{figure}
\begin{figure*}
\epsscale{1.0}
\plotone{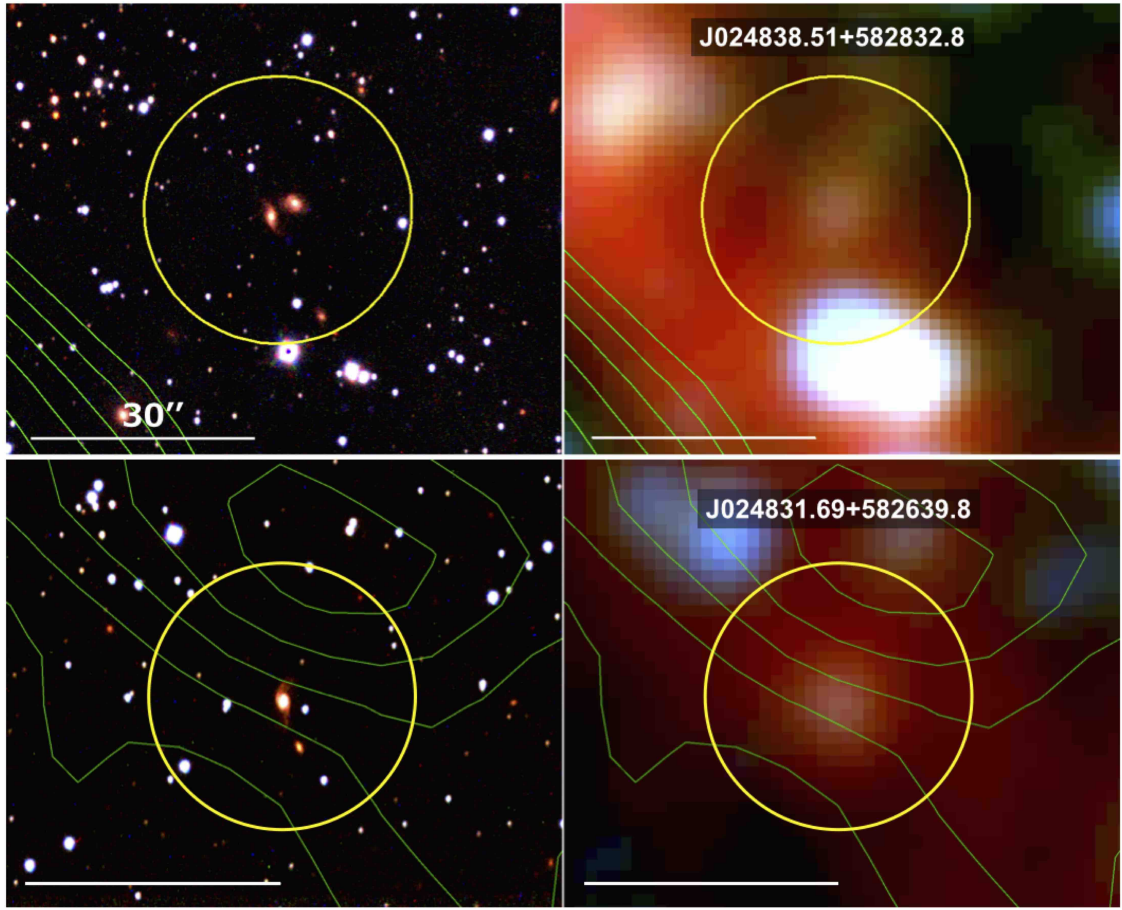}
\caption{
Subaru $JHK_{S}$ (left) and {\it WISE} 3.4, 4.6, and 12 {\micron} (right) pseudo-color images demonstrating contamination by three galaxies (possible AGNs) in the CO core of Cloud 2.
Green contours show the $^{12}$CO (1-0) distribution from the NRO 45 m telescope with contour levels of 3$\sigma$, 5$\sigma$, 7$\sigma$, 9$\sigma$, and 11$\sigma$.
The yellow circle in each panel shows the location of a candidate star-forming region, which turned out to be a galaxy in the corresponding Subaru NIR image
(top: $l$ = 137$^\circ$.760, $b$ = $-$0$^\circ$.981; bottom: $l$ = 137$^\circ$.761, $b$ = $-$1$^\circ$.015).
 }  
\label{DC2_Con} 
\end{figure*}
\begin{figure*}
\epsscale{1.0}
\plotone{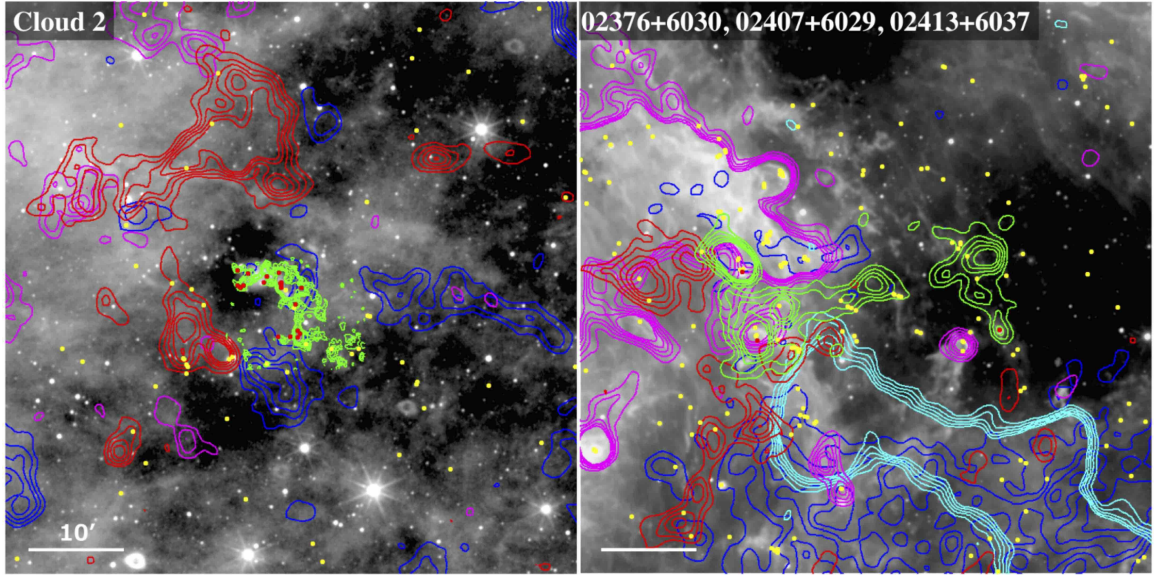}
\caption{
Left: distribution of foreground molecular clouds in and around Digel Cloud 2.
The gray-scale background shows the {\it WISE} 12 {\micron} images.
Green contours show the $^{12}$CO(1-0) distribution from the NRO 45 m telescope with contour levels of 3$\sigma$, 5$\sigma$, 7$\sigma$, 9$\sigma$, and 11$\sigma$, the same as in Figure \ref{CO-NIR-MIR-DC12}.
Red, magenta, and blue contours show the $^{12}$CO(1-0) distribution of foreground molecular clouds from the FCRAO data with contour levels of 3$\sigma$, 5$\sigma$, 7$\sigma$, 9$\sigma$, and 11$\sigma$
red: $v_{\rm LSR}$ = $-$43.5 to $-$47.6 km s$^{-1}$, 1 $\sigma$ = 0.28 K km s$^{-1}$;
magenta: $v_{\rm LSR}$ = $-$35.3 to $-$38.6 km s$^{-1}$, 1 $\sigma$ = 0.25 K km s$^{-1}$;
blue: $v_{\rm LSR}$ = $-$0.6 to $-$8.9 km s$^{-1}$, 1 $\sigma$ = 0.42 K km s$^{-1}$).
Right: distribution of the foreground molecular clouds in and around the 02376+6030, 02407+6029, and 02413+6037 clouds.
The gray-scale background shows the {\it WISE} 12 {\micron} images.
Green contours show the $^{12}$CO(1-0) distribution from the FCRAO data with contour levels of 3$\sigma$, 5$\sigma$, 7$\sigma$, 9$\sigma$, and 11$\sigma$, the same as in Figure \ref{MIR-SnellSFRs}.
Red, magenta, cyan, and blue contours show the $^{12}$CO(1-0) distribution of the foreground molecular clouds from the FCRAO data with contour levels of 3$\sigma$, 5$\sigma$, 7$\sigma$, 9$\sigma$, and 11$\sigma$,
(red: $v_{\rm LSR}$ = $-$49.3 to $-$55.9 km s$^{-1}$, 1 $\sigma$ = 0.37 K km s$^{-1}$;
magenta: $v_{\rm LSR}$ = $-$32.0 to $-$46.0 km s$^{-1}$, 1 $\sigma$ = 0.56 K km s$^{-1}$;
cyan: $v_{\rm LSR}$ = $-$8.1 to $-$16.3 km s$^{-1}$, 1 $\sigma$ = 0.42 K km s$^{-1}$;
blue: $v_{\rm LSR}$ = 5.1 to $-$3.1 km s$^{-1}$, 1 $\sigma$ = 0.42 K km s$^{-1}$).
}  
\label{Foreground-clouds} 
\end{figure*}
\begin{figure*}
\epsscale{1.0}
\plotone{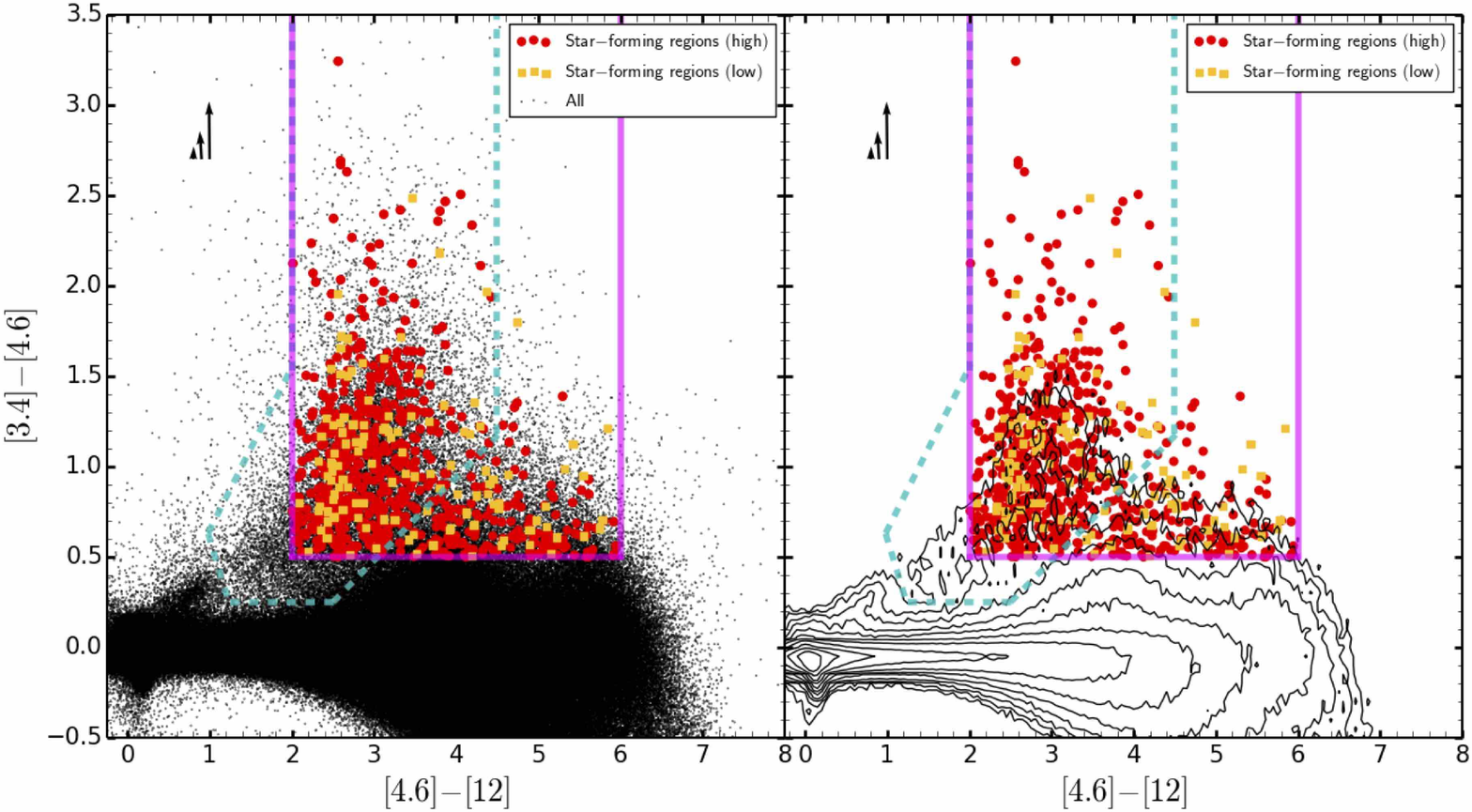}
\caption{
The [3.4]$-$[4.6] vs. [3.6]$-$[12] color-color diagram of all the AllWISE catalog sources in the FCRAO CO survey of the outer Galaxy area 
(102$^\circ$.49 $\le$ $l$ $\le$ 141$^\circ$.54 and $-$3$^\circ$.03 $\le$ $b$ $\le$ 5$^\circ$.41).
We selected the sources that do not have {\it cc\_flags} in all of the four bands and have S/N $>$ 5 in all of the 3.4, 4.6, and 12 {\micron} bands (926,132 sources in the area).
Red circles and yellow squares show the candidate star-forming regions with the contamination rate of their parental clouds of $<$ 30 \% and $\ge$ 30 \%, respectively.
The black dots in the left panel show all the sources in the area.
The black contours in the right panel show the distributions of all the sources in the area (10, 20, 40, 80, 160, 320, 640, 1280, and 2560 independent data points per 0.05 cell).
The black arrows show the extinction 
vectors for, from left to right, $A_{K}$ = 0.4, 0.8, and 2 \citep{McClure09,Koenig14}.
The polygons delineated with cyan dashes show the region of individual YSOs from \citet{Koenig14}.
The magenta box shows our defined region of distant star-forming regions.
}
\label{CC_FCRAO}
\end{figure*}
\begin{figure}
\epsscale{1.0}
\plotone{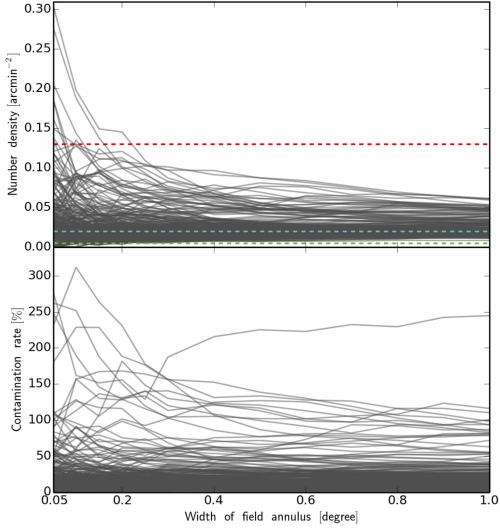}
\caption{
Top: number density of candidate star-forming regions detected as a function of the width of the annulus for the trial region in the field region around clouds with associated candidate star-forming regions.
The red and cyan dotted lines show the mean number density of candidate star-forming regions in the cloud (0.13 arcmin$^{-2}$) and field regions (0.020 arcmin$^{-2}$), respectively.
The green dotted line shows the total number density of QSOs,  AGNs, and PNe (0.0045 arcmin$^{-2}$; see Section \ref{sec:3.5}).
Bottom: contamination rate as a function of width, the same as in the top panel.
}  
\label{ND-CR} 
\end{figure}
\begin{figure}
\epsscale{0.9}
\plotone{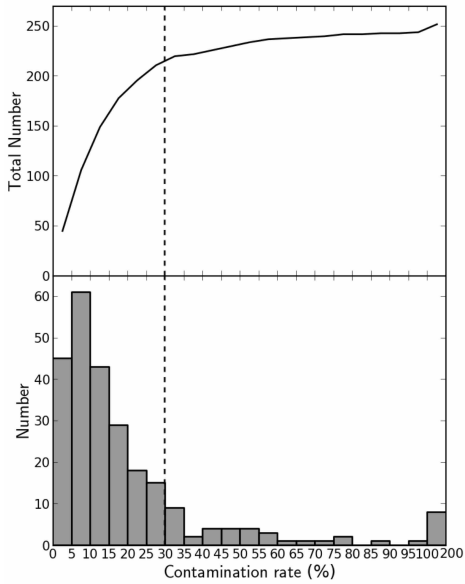}
\caption{
Histogram and plot of the contamination rates for all the 252 molecular clouds with associated candidate star-forming regions.
Top: cumulative number of the molecular clouds.
Bottom: number of the molecular clouds with a bin size of contamination rate of 5\%.
The black dotted line shows the contamination threshold of 30\%.
Clouds with a contamination rate lower than 30\% (211 clouds) are regarded as high-reliability candidates.
}  
\label{Contami} 
\end{figure}
\begin{figure*}
\epsscale{1.0}
\plotone{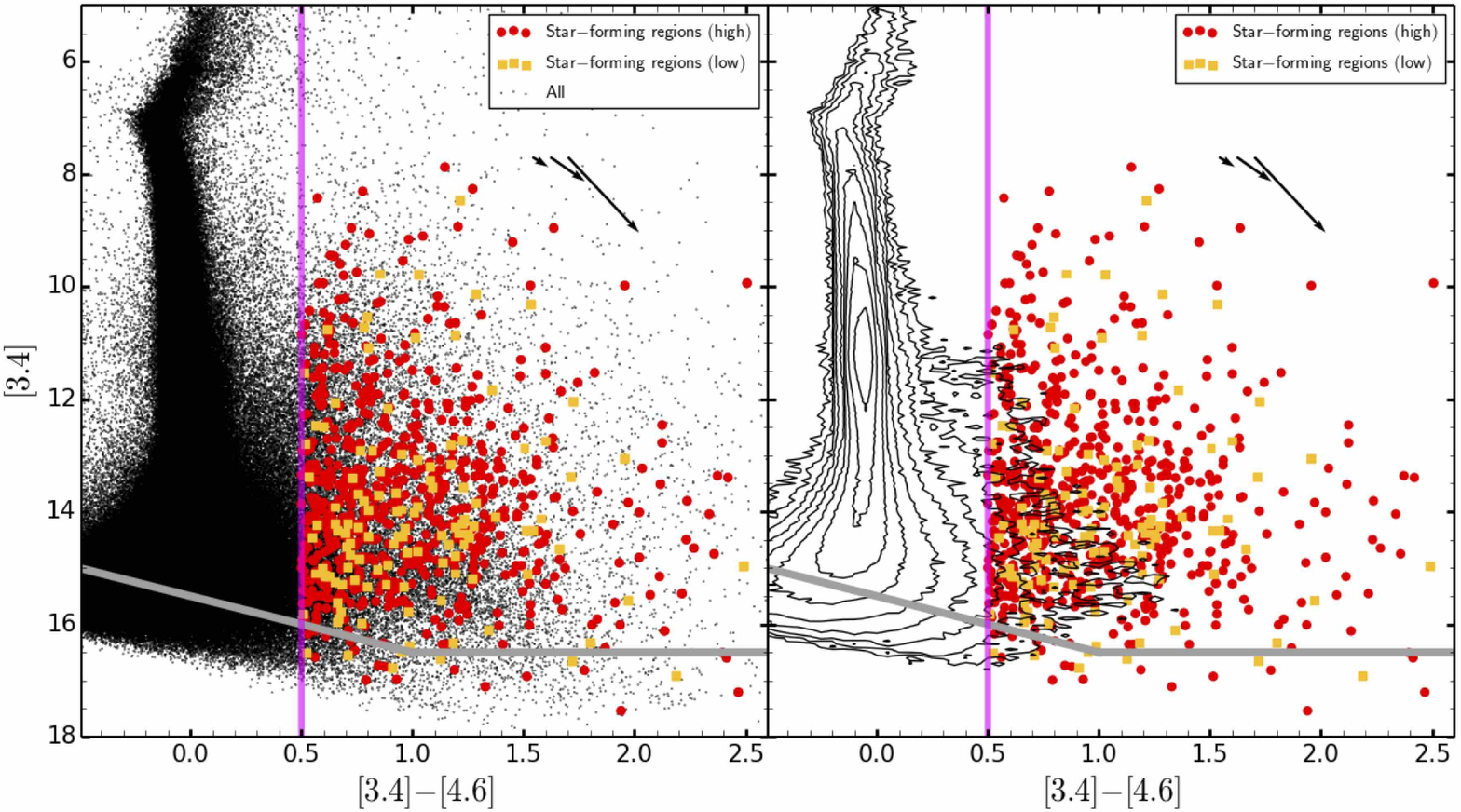}
\caption{
The [3.4] vs. [3.4]$-$[4.6] color-magnitude diagram of all of the AllWISE catalog sources in the FCRAO CO survey of the outer Galaxy area (102$^\circ$.49 $\le$ $l$ $\le$ 141$^\circ$.54 and $-$3$^\circ$.03 $\le$ $b$ $\le$ 5$^\circ$.41).
We selected the sources that do not have {\it cc\_flags} in all of the four bands and have S/N $>$ 5 in all of the 3.4, 4.6, and 12 {\micron} bands (926,132 sources in the area).
Red circles and yellow squares show the candidate star-forming regions with a contamination rate of their parental clouds of $<$ 30 \% and $\ge$ 30 \%, respectively.
The black dots in the left panel show all the sources in the area.
The black contours in the right panel show the distribution of all the sources in the area (10, 20, 40, 80, 160, 320, 640, 1280, and 2560 independent data points per 0.05 cell).
Gray lines show the average detection limit for the minimum integration for eight frames \citep[16.5 mag for 3.4 $\mu$m and 15.5 mag for 4.6 $\mu$m;][]{Wright10}.
The black arrows show the extinction vectors for, from left to right, $A_{K}$ = 0.4, 0.8, and 2 \citep{McClure09,Koenig14}.
The magenta lines show our defined boundary for distant star-forming regions. 
} 
\label{CM_FCRAO}
\end{figure*}
\begin{figure*}
\epsscale{1.2}
\plotone{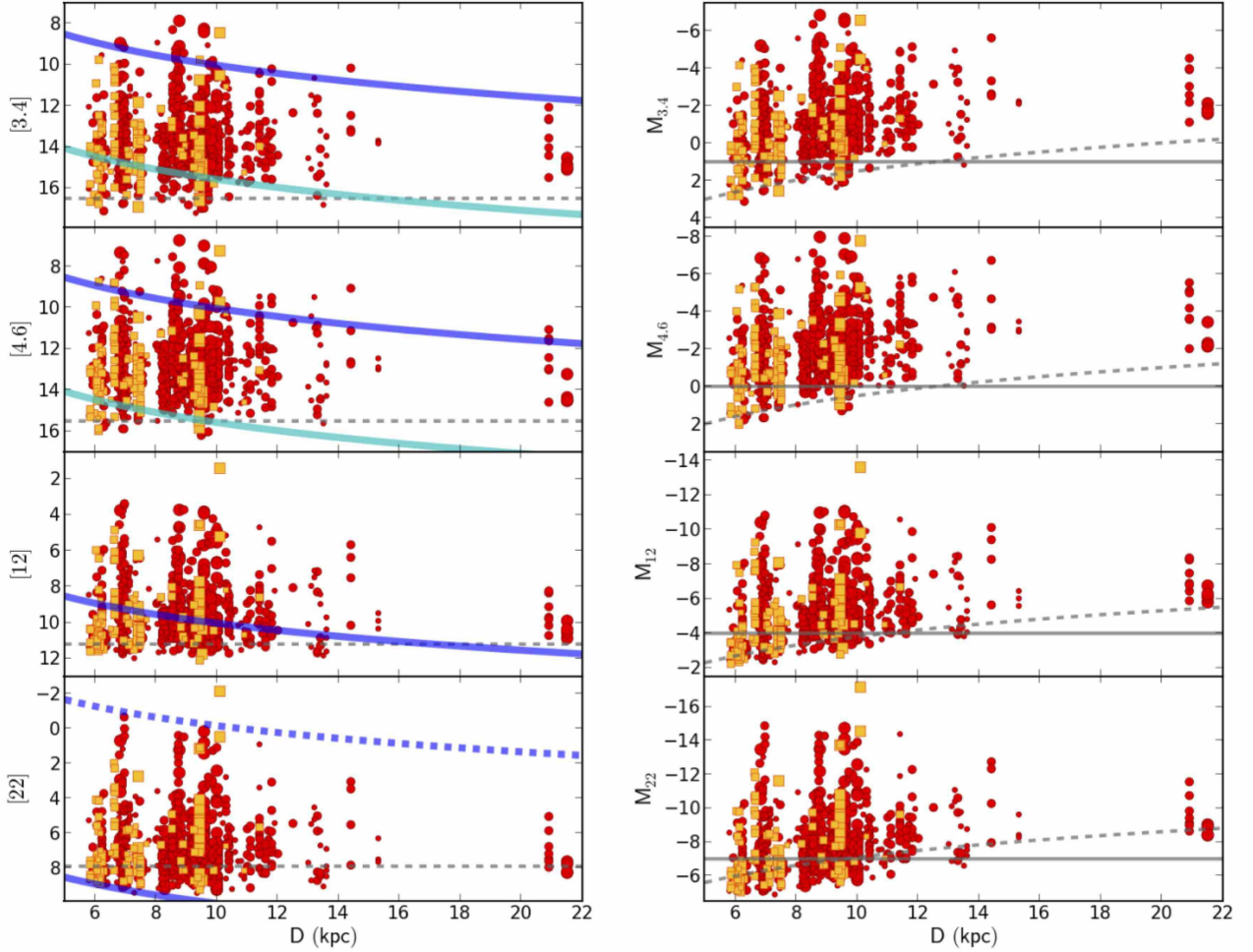}
\caption{
Left: apparent magnitude variation as a function of the kinematic distance for the candidate star-forming regions.
Red circles and yellow squares show the candidate star-forming regions with a contamination rate of their parental clouds of $<$ 30\% and $\ge$ 30\%, respectively.
The size of the symbols indicates the mass of their parental clouds 
(small: 10$^2$ $M_\odot$ $\le$ $M_{\rm cloud}$ $<$ 10$^3$ $M_\odot$; middle: 10$^3$ $M_\odot$ $\le$ $M_{\rm cloud}$ $<$ 10$^4$ $M_\odot$; large: 10$^4$ $M_\odot$ $\le$ $M_{\rm cloud}$).
Note that 11 candidates with kinematic distances of more than 20 kpc ($R_{\rm G}$  $>$ 27 kpc) have already been found in previous study (sample star forming regions associated with Digel Cloud 2)
and are known to be actually located at a distance of 12 kpc ($R_{\rm G}$  = 19 kpc) from high-resolution optical spectra \citep[e.g.][]{Smartt96, Kobayashi08}.
The gray dotted lines show the average detection limit for the minimum integration for eight frames \citep[16.5, 15.5, 11.2, and 7.9 mag for 3.4, 4.6, 12, and 22 {\micron}, respectively;][]{Wright10}.
Cyan and blue curves show the apparent magnitude for the A0 and B0 stars in the main sequence \citep{Cox00}.
The blue dotted curve shows the 22 {\micron} apparent magnitude for the \ion{H}{2} regions ionized by B0 stars \citep{Anderson14}.
Right:  same as the left panel but for the absolute magnitudes.
The gray solid lines show the completeness limit up to $D$ $\sim$ 14 kpc.
} 
\label{Mag-D}
\end{figure*}
\begin{figure}
\epsscale{1.2}
\plotone{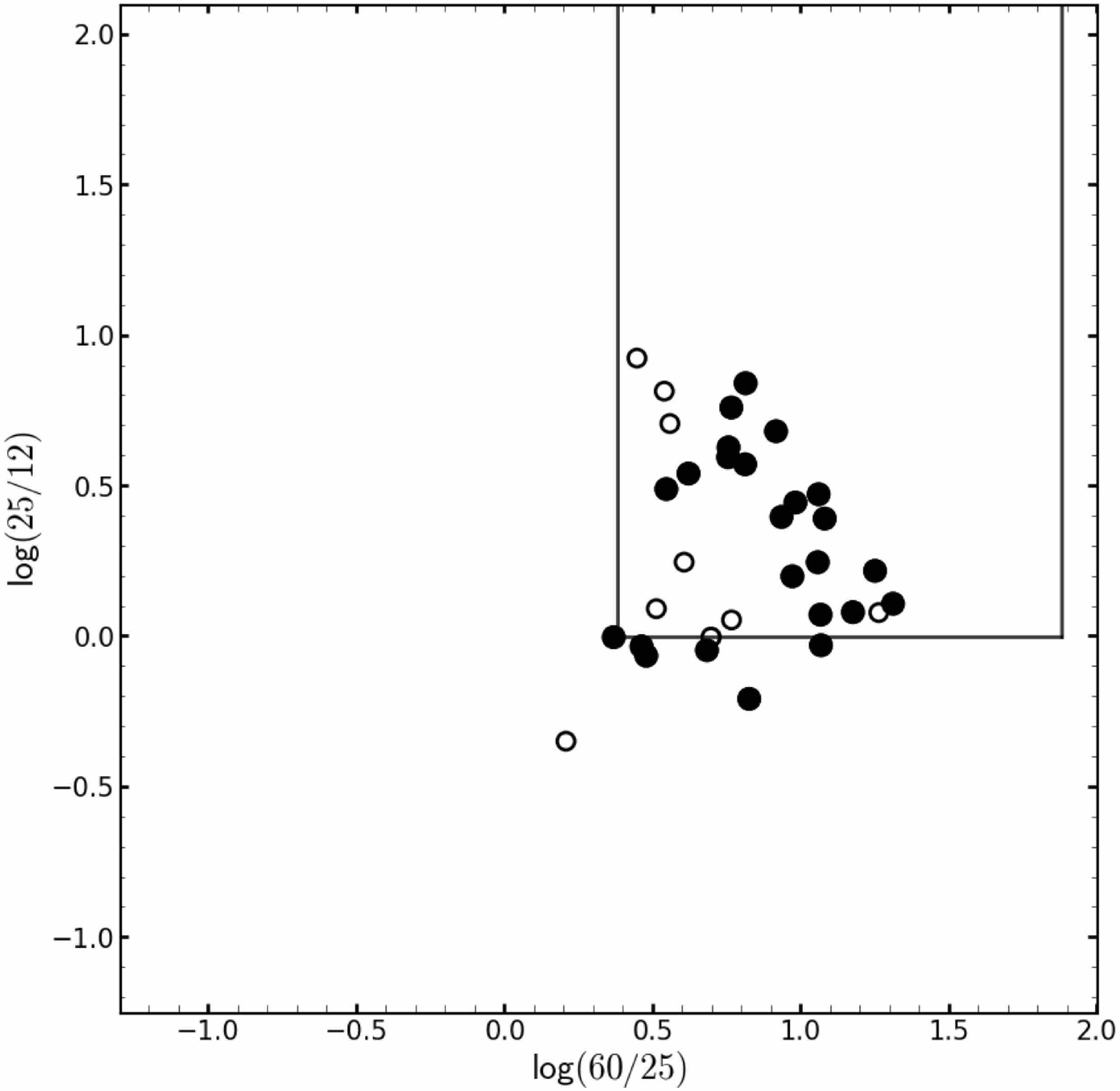}
\caption{
{\it IRAS} color-color diagram for the {\it IRAS} sources associated with the BKP clouds with $N_E$ $\ge$ 10$^{-2.0}$ at $R_{\rm G}$ $\ge$ 13.5 kpc identified by \citet{Kerton03}.
The filled and open circles indicate the {\it IRAS} sources detected and not detected, respectively, 
as candidate star-forming regions in our survey with the {\it WISE} data.
The black box indicates the region of representative star-forming regions by \citet{Wouterloot89}: $\log{(25 \micron/12 \micron)}$ $>$ 0, 0.38 $<$ $\log{(60 \micron/25 \micron)}$ $<$ 1.88.
}
\label{IRAS-CC}
\end{figure}
\clearpage
\begin{deluxetable*}{clccccccccc}
\tabletypesize{\normalsize}
\tablewidth{0pt}
\tablecaption{Molecular Clouds with Associated Sample Star-forming Regions in the FOG and EOG\label{tbl:sampleMClist}}
\tablehead{
\colhead{Region} &\colhead{Molecular Cloud} &\multicolumn{2}{c}{Galactic Coordinates}&
\colhead{$V_{\rm LSR}$} &\colhead{$D$} &\colhead{$R_{\rm G}$} &\colhead{Age}& \colhead{References}&\\
\colhead{} & \colhead{} & \colhead{$l$ (deg)} &\colhead{$b$ (deg)}&
\colhead{(km s$^{-1}$)} &\colhead{(kpc)} &\colhead{(kpc)} &\colhead{(Myr)}  &\colhead{}  &\colhead{} }
\startdata
EOG &Digel Cloud 1                       &131$^{\circ}.05$&1$^{\circ}.45$               &-101.8   &16    &22             &$<$ 1       &(1),(2)            &\\                            
        &Digel Cloud 2                       &137$^{\circ}.75$&-1$^{\circ}.00$              &-102.4  &12     &19             &0.5 -- 1.0   &(1),(3),(4)         &\\
        &WB89-789 cloud                   &195$^{\circ}.82$&-0$^{\circ}.57$              &34.01    &11.9  &20.2          & \nodata    &(5),(6)            &\\ \hline
FOG &01537+6154 cloud                &130$^{\circ}.539$&0$^{\circ}.263$           &-62.0     &6.5   &13.6           & \nodata    &(7),(8)            &\\     
        &01587+6148 cloud                 &131$^{\circ}.145$&0$^{\circ}.312$           &-61.0     &6.3   &13.5           & \nodata    &(7),(8)            &\\        
        &02071+6235 cloud                 &131$^{\circ}.856$&1$^{\circ}.332$           &-78.3     &9.4   &16.4           & \nodata    &(7),(8)            &\\   
        &02376+6030 cloud                 &135$^{\circ}.988$&0$^{\circ}.672$           &-78.1     &10.2 &17.4           & \nodata    &(7),(8)            &\\   
        &02383+6241 cloud                 &135$^{\circ}.182$&2$^{\circ}.694$           &-71.8     &8.5   &15.7           & \nodata    &(7),(8)            &\\   
        &02395+6244 cloud                 &135$^{\circ}.278$&2$^{\circ}.797$           &-71.6     &8.8   &16.0           & \nodata    &(7),(8)            &\\   
        &02407+6029 cloud                &136$^{\circ}.347$&0$^{\circ}.817$           &-74.4     &9.4   &16.8           & \nodata    &(7),(8)           &\\   
        &02413+6037 cloud                  &136$^{\circ}.357$&0$^{\circ}.958$           &-61.6     &6.9   &14.3           & \nodata    &(7),(8)            &\\ 
        &02421+6233 cloud                  &135$^{\circ}.627$&2$^{\circ}.765$           &-72.4     &8.9   &16.1           & \nodata    &(7),(8)            &\\   
        &02598+6008  cloud                 &138$^{\circ}.618$&1$^{\circ}.562$           &-59.6     &7.0   &14.5           & \nodata    &(7),(8)            &
\enddata
\tablenotetext{}{References. (1) \citet{Digel94}; (2) \citet{Izumi14}; (3) \citet{Kobayashi08}; (4) \citet{Yasui08}; 
(5) \citet{Brand94}; (6) \citet{Brand07}; (7) \citet{Heyer98}; (8) \citet{Snell02}}
\end{deluxetable*}

\begin{deluxetable*}{ccccccccccccccc}
\tabletypesize{\tiny}
\tablewidth{0pt}
\tablecaption{Sample and Candidate Star-forming Regions and the {\it WISE} Sources in the Outer Galaxy\label{tbl:sampleSFR}}
\tablehead{
\colhead{Region}&\colhead{Molecular}&\colhead{Star-forming $^a$} &\colhead{Type $^b$} &\colhead{AllWISE Source}&\multicolumn{2}{c}{Coordinate}&
\colhead{3.4 $\micron$}   &\colhead{$\sigma_{3.4}$}   &\colhead{4.6 $\micron$}   &\colhead{$\sigma_{4.6}$}&
\colhead{12 $\micron$}    &\colhead{$\sigma_{12}$}    &\colhead{22 $\micron$}    &\colhead{$\sigma_{22}$} \\
\colhead{}&\colhead{Cloud} &\colhead{Region}&                     &&\colhead{$l$}&\colhead{$b$}&
\colhead{(mag)}                &\colhead{(mag)}                  &\colhead{(mag)}               &\colhead{(mag)}        &
\colhead{(mag)}                &\colhead{(mag)}                  &\colhead{(mag)}               &\colhead{(mag)}  }
\startdata
EOG&Digel Cloud 1 &CAN               &-                         & J020411.94+631135.7 & 131.028 &  1.471 & 15.328 & 0.044 & 13.679 & 0.035 & 10.939 & 0.103 & 8.476 & 0.354  \\ 
&                      &Q                   &A                        & J020429.54+631412.8 & 131.047 &  1.522 & 13.059 & 0.025 & 11.901 & 0.022 & 9.156 & 0.035 & 6.704 & 0.074  \\ 
&                      &Cloud 1a         &EC                      & J020417.32+631418.9 & 131.025 &  1.517 & 14.053 & 0.027 & 13.195 & 0.029 & 10.718 & 0.096 & 7.885 & 0.223  \\ 
&                      &Cloud 1a         &EC                      & J020418.20+631436.0 & 131.025 &  1.522 & 14.310 & 0.037 & 13.559 & 0.044 & 10.948 & 0.142 & 7.711 & 0.270  \\ 
&                      &Cloud 1a         &EC                      & J020417.77+631441.6 & 131.024 &  1.524 & 14.444 & 0.040 & 13.592 & 0.046 & 11.160 & 0.174 & 8.082 & null  \\ 
&                      &Cloud 1b         &EC                      & J020508.31+630452.9 & 131.161 &  1.393 & 13.781 & 0.030 & 12.462 & 0.027 & 9.475 & 0.044 & 6.297 & 0.065  \\ 
&                      &Cloud 1b         &EC                      & J020508.25+630511.5 & 131.160 &  1.398 & 13.692 & 0.026 & 12.997 & 0.029 & 9.904 & 0.050 & 7.505 & 0.139  \\ 
&                     &Q                   &A                        & J020504.94+630314.9 & 131.163 &  1.365 & 13.830 & 0.030 & 12.895 & 0.028 & 10.340 & 0.069 & 7.650 & 0.168  \\ 
&Digel Cloud 2  & Cloud 2-N     &EC                       & J024842.33+582847.2 & 137.766 & -0.973 & 13.997 & 0.030 & 13.118 & 0.029 & 9.601 & 0.048 & 6.757 & 0.078  \\ 
&                      &Q                   &A                         & J024912.15+582901.6 & 137.823 & -0.941 & 15.288 & 0.049 & 14.764 & 0.073 & 10.592 & 0.119 & 8.000 & 0.225  \\ 
&                      &Q                   &A                         & J024915.44+582848.2 & 137.831 & -0.942 & 15.004 & 0.051 & 14.147 & 0.046 & 10.285 & 0.082 & 7.091 & 0.113  \\ 
&                      &Q                   &A                         & J024917.52+582900.8 & 137.833 & -0.936 & 15.535 & 0.042 & 14.788 & 0.058 & 11.284 & 0.172 & 8.280 & null  \\ 
&                      &Q                   &A                         & J024914.79+582845.1 & 137.830 & -0.943 & 15.225 & 0.050 & 13.561 & 0.035 & 9.540 & 0.061 & 6.719 & 0.094  \\ 
&                      &Q                   &A                         & J024853.48+582954.8 & 137.780 & -0.946 & 13.860 & 0.034 & 13.233 & 0.030 & 9.869 & 0.065 & 7.285 & 0.125  \\ 
&                      &Q                   &A                         & J024908.91+583014.5 & 137.807 & -0.926 & 15.431 & 0.047 & 14.462 & 0.050 & 10.243 & 0.103 & 7.646 & 0.230  \\ 
&                      &IRS 1               &A                         & J024856.41+582919.7 & 137.790 & -0.952 & 11.713 & 0.023 & 11.001 & 0.020 & 7.374 & 0.018 & 5.365 & 0.050  \\ 
&                      &Q                   &A                         & J024917.38+583032.6 & 137.822 & -0.914 & 13.729 & 0.028 & 13.126 & 0.030 & 10.539 & 0.130 & 8.390 & 0.328  \\ 
&                      &Cloud 2-N       &EC                       & J024842.95+582912.4 & 137.764 & -0.966 & 15.546 & 0.055 & 14.646 & 0.063 & 10.554 & 0.127 & 7.539 & 0.146  \\ 
&                      &Cloud 2-N       &EC                       & J024841.97+582916.7 & 137.762 & -0.966 & 14.707 & 0.037 & 14.090 & 0.045 & 11.071 & 0.194 & 8.718 & 0.423  \\ 
&                      &Cloud 2-N       &EC                       & J024842.40+582904.2 & 137.764 & -0.969 & 15.278 & 0.049 & 14.404 & 0.053 & 10.539 & 0.120 & 7.537 & 0.158  \\ 
&                      &Q                   &A                         & J024825.92+582846.8 & 137.734 & -0.989 & 14.539 & 0.033 & 13.228 & 0.030 & 9.903 & 0.056 & 7.629 & 0.139  \\ 
&                      &Q                   &A                         & J024843.94+583021.8 & 137.758 & -0.948 & 14.190 & 0.030 & 12.980 & 0.028 & 10.328 & 0.087 & 7.501 & 0.128  \\ 
&                      &CAN               & \nodata                         & J024801.66+582222.8 & 137.732 & -1.108 & 14.423 & 0.030 & 13.028 & 0.028 & 9.707 & 0.042 & 7.472 & 0.129  \\ 
&                      &CAN               &\nodata                         & J024853.76+582046.1 & 137.847 & -1.083 & 13.742 & 0.042 & 12.739 & 0.036 & 9.156 & 0.068 & 6.975 & 0.152  \\ 
&                      &Q                   &A                         & J024830.64+582700.5 & 137.756 & -1.011 & 14.859 & 0.040 & 14.356 & 0.050 & 10.871 & 0.110 & 8.271 & 0.222  \\ 
&                      &IRS 3               &A                         & J024826.90+582357.6 & 137.771 & -1.060 & 13.571 & 0.033 & 12.424 & 0.028 & 9.812 & 0.061 & 7.182 & 0.155  \\ 
&                      &Cloud 2-S        &EC                       & J024828.69+582331.9 & 137.777 & -1.065 & 12.686 & 0.023 & 11.495 & 0.021 & 8.358 & 0.026 & 5.042 & 0.034  \\ 
&                      &IRS 5               &A                         & J024844.84+582336.1 & 137.809 & -1.049 & 12.076 & 0.023 & 11.062 & 0.020 & 8.240 & 0.022 & 5.856 & 0.046  \\ 
&                      &IRS 4               &A                         & J024835.25+582336.1 & 137.790 & -1.058 & 12.626 & 0.025 & 11.611 & 0.022 & 9.124 & 0.038 & 6.767 & 0.085  \\ 
&                      &Q                   &A                         & J024829.02+582414.5 & 137.773 & -1.054 & 15.489 & 0.068 & 14.590 & 0.052 & 10.719 & 0.110 & 7.624 & 0.133  \\ 
&                      &CAN               & \nodata                         & J024822.19+582249.7 & 137.770 & -1.082 & 14.037 & 0.028 & 12.966 & 0.027 & 10.166 & 0.063 & 7.960 & 0.214  \\ 
&WB89-789 cloud    & \nodata                     &EC                       & J061724.10+145431.6 & 195.823 & -0.569 & 10.588 & 0.020 & 9.948 &  0.020 & 4.746 &  0.011 & 1.834 &  0.031  \\
&                      & \nodata                    &EC                       & J061724.02+145440.7 & 195.821 & -0.568 & 10.313 &  0.026 & 8.804 &  0.020 & 4.944 &  0.020 & 1.706 &  0.027  \\
&                      & \nodata                     &EC                       & J061725.21+145449.6 & 195.821 & -0.563 & 11.268 &  0.058 & 10.231 &  0.028 & 5.045 &  0.015 & 1.924 &  0.020  \\ \hline      
FOG&01537+6154 cloud   &CAN                & \nodata                         & J015724.02+620703.1 & 130.560 &  0.227 & 9.537 & 0.023 & 8.333 & 0.021 & 5.166 & 0.016 & 2.981 & 0.023  \\
&                      &01537+6154     &EC                       & J015719.28+620914.7 & 130.542 &  0.260 & 11.764 & 0.023 & 11.203 & 0.022 & 9.205 & 0.037 & 4.560 & 0.034  \\ 
&                      &CAN                & \nodata                         & J015718.83+620832.0 & 130.544 &  0.248 & 14.399 & 0.029 & 13.758 & 0.032 & 11.294 & 0.185 & 8.089 & 0.249  \\ 
&                      &01537+6154     &EC                       & J015718.61+620931.3 & 130.540 &  0.264 & 9.801 & 0.023 & 7.651 & 0.020 & 5.538 & 0.013 & 2.729 & 0.016  \\ 
&                      &CAN                & \nodata                         & J015721.61+621033.4 & 130.541 &  0.282 & 15.445 & 0.041 & 13.453 & 0.029 & 9.897 & 0.046 & 6.550 & 0.063  \\ 
&                      &CAN                & \nodata                         & J015736.47+620554.3 & 130.589 &  0.214 & 13.845 & 0.028 & 13.288 & 0.029 & 10.966 & 0.142 & 8.062 & 0.221  \\ 
&                      &CAN                & \nodata                         & J015727.30+620644.2 & 130.568 &  0.223 & 13.537 & 0.028 & 12.468 & 0.028 & 9.598 & 0.049 & 6.504 & 0.076  \\ 
&                      &CAN                & \nodata                        & J015741.70+620511.2 & 130.602 &  0.205 & 12.432 & 0.024 & 11.651 & 0.022 & 9.357 & 0.037 & 7.128 & 0.099  \\ 
&                      &CAN                & \nodata                         & J015738.91+620533.1 & 130.595 &  0.210 & 12.321 & 0.048 & 11.720 & 0.022 & 9.619 & 0.047 & 7.318 & 0.126  \\ 
&                      &CAN                & \nodata                         & J015748.03+620503.9 & 130.614 &  0.207 & 9.818 & 0.022 & 9.072 & 0.021 & 6.861 & 0.017 & 5.003 & 0.033  \\ 
&01587+6148 cloud   &CAN                & \nodata                         & J020239.51+620525.3 & 131.161 &  0.362 & 14.789 & 0.033 & 13.923 & 0.038 & 11.324 & 0.173 & 8.290 & 0.282  \\ 
&                      &CAN                & \nodata                         & J020238.68+620543.8 & 131.158 &  0.367 & 15.452 & 0.070 & 14.775 & 0.071 & 11.362 & 0.184 & 8.130 & 0.233  \\ 
&                      &CAN                & \nodata                        & J020236.09+620513.8 & 131.156 &  0.357 & 17.110 & 0.118 & 15.782 & 0.115 & 11.072 & 0.142 & 7.114 & 0.103  \\ 
&                      &01587+6148     &EC                       & J020223.40+620245.3 & 131.143 &  0.311 & 11.472 & 0.023 & 10.891 & 0.021 & 5.935 & 0.014 & 3.726 & 0.020  \\ 
&02071+6235 cloud   &02071+6235      &EC w/ \ion{H}{2}  & J021049.90+624910.4 & 131.857 & 1.333  & 9.774  &  0.020 & 8.923 &  0.015 & 4.537 &  0.014 & 1.098 &  0.013 \\
&02376+6030 cloud   &02376+6030      &EC                       & J024129.21+604327.8 & 135.988 &  0.673 & 10.677 & 0.022 & 10.159 & 0.020 & 5.148 & 0.015 & 2.779 & 0.018  \\ 
&                      &CAN                & \nodata                         & J024156.78+605113.4 & 135.986 &  0.814 & 15.289 & 0.051 & 14.023 & 0.044 & 10.714 & 0.134 & 8.412 & 0.336  \\ 
&                      &CAN                & \nodata                         & J024202.34+605240.4 & 135.986 &  0.841 & 15.990 & 0.094 & 15.428 & 0.107 & 10.233 & 0.116 & 8.264 & 0.279  \\ 
&                      &CAN                & \nodata                         & J024202.89+605232.2 & 135.988 &  0.839 & 16.101 & 0.105 & 15.390 & 0.110 & 10.490 & 0.143 & 8.193 & 0.263  \\ 
&                      &CAN                & \nodata                         & J024204.42+605202.2 & 135.995 &  0.833 & 16.339 & 0.191 & 15.470 & 0.115 & 9.944 & 0.117 & 8.167 & 0.250  \\ 
&                      &CAN                & \nodata                         & J024148.47+604853.0 & 135.987 &  0.772 & 15.820 & 0.052 & 14.573 & 0.054 & 11.229 & 0.162 & 8.374 & null  \\ 
&02383+6241 cloud    &CAN                & \nodata                         & J024241.28+625436.3 & 135.213 &  2.722 & 13.539 & 0.027 & 12.843 & 0.030 & 8.172 & 0.040 & 5.197 & 0.042  \\ 
&                      &CAN                & \nodata                         & J024218.97+625405.0 & 135.178 &  2.697 & 10.779 & 0.027 & 10.148 & 0.024 & 4.700 & 0.015 & 1.179 & 0.034  \\ 
&                      &CAN                & \nodata                         & J024224.73+625407.2 & 135.188 &  2.702 & 11.196 & 0.021 & 10.637 & 0.021 & 5.297 & 0.012 & 1.596 & 0.020  \\ 
&                      &CAN                & \nodata                         & J024220.30+625447.2 & 135.176 &  2.708 & 13.346 & 0.026 & 12.741 & 0.025 & 7.580 & 0.018 & 3.090 & 0.008  \\ 
&                      &CAN                & \nodata                         & J024219.76+625430.3 & 135.177 &  2.704 & 11.636 & 0.016 & 11.042 & 0.014 & 5.957 & 0.006 & 1.138 & 0.004  \\ 
&                      &CAN                & \nodata                         & J024222.00+625438.4 & 135.180 &  2.707 & 13.455 & 0.019 & 12.914 & 0.026 & 7.249 & 0.015 & 1.960 & 0.006  \\ 
&                      &02383+6241a    &EC w/ \ion{H}{2}  & J024221.73+625403.6 & 135.183 &  2.698 & 9.942 & 0.025 & 9.344 & 0.021 & 4.601 & 0.016 & 0.384 & 0.018  \\ 
&                      &CAN                & \nodata                         & J024219.87+625404.0 & 135.180 &  2.697 & 10.650 & 0.029 & 9.917 & 0.027 & 4.705 & 0.019 & 0.871 & 0.029  \\ 
&                      &CAN                & \nodata                         & J024238.04+625430.8 & 135.208 &  2.718 & 12.918 & 0.034 & 12.414 & 0.037 & 7.502 & 0.026 & 4.628 & 0.032  \\ 
&                      &02383+6241b    &EC                       & J024239.14+625425.3 & 135.211 &  2.718 & 12.029 & 0.028 & 10.790 & 0.022 & 7.045 & 0.022 & 4.247 & 0.037  \\ 
&                      &CAN                & \nodata                         & J024216.31+625322.9 & 135.179 &  2.684 & 13.519 & 0.016 & 12.852 & 0.014 & 7.245 & 0.005 & 4.223 & 0.018  \\ 
&                      &CAN                & \nodata                        & J024217.33+625324.8 & 135.180 &  2.685 & 13.283 & 0.026 & 12.582 & 0.025 & 6.782 & 0.014 & 3.904 & 0.021  \\ 
&02407+6029 cloud  &CAN                & \nodata                         & J024459.59+604345.7 & 136.376 &  0.857 & 13.900 & 0.035 & 12.879 & 0.032 & 9.912 & 0.083 & 6.904 & 0.090  \\ 
&                      &CAN                & \nodata                         & J024422.18+604114.4 & 136.324 &  0.786 & 14.434 & 0.089 & 13.479 & 0.139 & 8.852 & 0.093 & 4.591 & 0.036  \\ 
&                      &CAN                & \nodata                         & J024422.36+604133.8 & 136.322 &  0.791 & 14.265 & 0.033 & 13.072 & 0.035 & 8.978 & 0.078 & 5.286 & 0.032  \\ 
&                      &CAN                & \nodata                         & J024421.32+604127.5 & 136.321 &  0.789 & 14.220 & 0.047 & 13.569 & 0.048 & 8.659 & 0.077 & 5.123 & 0.045  \\ 
&                      &CAN                & \nodata                         & J024425.77+604143.5 & 136.328 &  0.797 & 14.457 & 0.160 & 13.205 & 0.139 & 9.536 & 0.202 & 6.264 & 0.155  \\ 
&                      &CAN                & \nodata                         & J024427.51+604136.2 & 136.332 &  0.796 & 14.218 & 0.088 & 12.995 & 0.070 & 8.704 & 0.060 & 5.711 & 0.099  \\ 
&                      &CAN                & \nodata                        & J024458.97+604250.7 & 136.381 &  0.842 & 12.726 & 0.026 & 11.505 & 0.026 & 8.368 & 0.029 & 5.431 & 0.049  \\ 
&                      &CAN                & \nodata                         & J024458.20+604259.7 & 136.379 &  0.844 & 13.688 & 0.037 & 12.936 & 0.049 & 8.816 & 0.033 & 5.776 & 0.052  \\ 
&                      &02407+6029b  &EC                        & J024457.48+604225.6 & 136.381 &  0.835 & 12.032 & 0.025 & 10.310 & 0.022 & 7.717 & 0.026 & 4.069 & 0.031  \\ 
&                      &CAN                & \nodata                         & J024440.19+604230.6 & 136.349 &  0.821 & 10.755 & 0.025 & 10.139 & 0.021 & 4.591 & 0.016 & 1.142 & 0.017  \\ 
&                      &CAN                & \nodata                         & J024332.92+604537.3 & 136.203 &  0.810 & 16.483 & 0.162 & 15.535 & 0.130 & 9.988 & 0.103 & 8.747 & 0.372  \\ 
&                      &CAN                & \nodata                         & J024259.82+604713.3 & 136.130 &  0.807 & 15.010 & 0.038 & 14.414 & 0.051 & 9.834 & 0.061 & 6.244 & 0.065  \\ 
&                      &CAN                & \nodata                         & J024300.53+604720.4 & 136.131 &  0.809 & 15.095 & 0.063 & 14.530 & 0.055 & 9.977 & 0.063 & 6.843 & 0.082  \\ 
&                      &CAN                & \nodata                         & J024253.79+604703.0 & 136.120 &  0.799 & 15.143 & 0.058 & 14.534 & 0.061 & 9.316 & 0.047 & 5.646 & 0.044  \\ 
&                      &CAN                & \nodata                         & J024256.29+604704.5 & 136.125 &  0.801 & 14.615 & 0.038 & 13.912 & 0.040 & 8.934 & 0.038 & 5.817 & 0.064  \\ 
&                      &CAN                & \nodata                         & J024438.68+604603.1 & 136.321 &  0.873 & 12.806 & 0.034 & 11.636 & 0.026 & 9.280 & 0.068 & 6.596 & 0.100  \\ 
&                      &CAN                & \nodata                         & J024336.65+604551.4 & 136.208 &  0.817 & 16.550 & 0.172 & 15.844 & 0.153 & 10.065 & 0.149 & 8.579 & null  \\ 
&                      &CAN                & \nodata                         & J024339.81+604601.4 & 136.213 &  0.822 & 16.615 & 0.168 & 15.492 & 0.123 & 10.071 & 0.113 & 8.329 & null  \\ 
&                      &CAN                & \nodata                         & J024346.10+604637.8 & 136.220 &  0.837 & 15.895 & 0.067 & 15.227 & 0.091 & 10.130 & 0.188 & 8.802 & 0.526  \\ 
&02413+6037 cloud   &CAN                & \nodata                         & J024512.35+604951.0 & 136.356 &  0.959 & 9.945 & 0.163 & 9.368 & 0.099 & 5.615 & 0.082 & 2.540 & 0.030  \\ 
&                      &CAN                & \nodata                         & J024514.20+605059.1 & 136.352 &  0.978 & 11.808 & 0.025 & 10.998 & 0.022 & 7.252 & 0.033 & 4.747 & 0.035  \\ 
&                      &CAN                & \nodata                         & J024514.44+605050.7 & 136.353 &  0.976 & 11.918 & 0.027 & 11.315 & 0.026 & 6.741 & 0.027 & 4.667 & 0.039  \\ 
&                      &CAN                & \nodata                         & J024545.93+605141.7 & 136.405 &  1.016 & 14.924 & 0.075 & 13.897 & 0.052 & 10.519 & 0.178 & 7.722 & 0.314  \\ 
&                      &CAN                & \nodata                         & J024527.40+605513.5 & 136.346 &  1.054 & 14.674 & 0.136 & 14.052 & 0.129 & 8.558 & 0.212 & 6.046 & 0.172  \\ 
&                      &CAN                & \nodata                         & J024510.24+604956.4 & 136.352 &  0.959 & 10.572 & 0.023 & 9.502 & 0.021 & 5.327 & 0.017 & 0.425 & 0.008  \\ 
&                      &02413+6037     &EC                       & J024510.79+604937.1 & 136.355 &  0.955 & 9.941 & 0.024 & 7.435 & 0.020 & 3.379 & 0.014 & -0.009 & 0.015  \\ 
&02421+6233 cloud    &02421+6233     &EC w/ \ion{H}{2}  & J024607.12+624630.3 & 135.626 & 2.765  & 9.447   &  0.022 & 8.817 &  0.020 & 3.721 &  0.014 & 0.474 &  0.016 \\
&                      &CAN                & \nodata                         & J024558.90+624708.0 & 135.607 & 2.767 & 12.086 &  0.023 & 11.316 &  0.021 & 7.805 &  0.021 & 5.131 &  0.036 \\
&02598+6008 cloud   & CAN                & \nodata                         & J030311.26+602007.6 & 138.548 &  1.521 & 12.065 & 0.024 & 11.302 & 0.024 & 7.219 & 0.023 & 5.658 & 0.041  \\ 
&                      &CAN                & \nodata                         & J030258.27+601920.8 & 138.531 &  1.497 & 11.484 & 0.024 & 10.787 & 0.021 & 7.898 & 0.061 & 6.088 & 0.113  \\ 
&                      &CAN                & \nodata                         & J030336.77+602055.1 & 138.588 &  1.558 & 15.035 & 0.050 & 11.792 & 0.023 & 9.228 & 0.117 & 5.419 & 0.078  \\ 
&                      &02598+6008     &EC                        & J030350.14+602013.2 & 138.618 &  1.562 & 9.211 & 0.023 & 7.760 & 0.020 & 4.075 & 0.014 & 1.167 & 0.011  \\
\enddata
\tablenotetext{a}{Q: star-forming regions (cluster/stellar aggregates) that are unpublished yet but identified separately from our Subaru and QUIRC data (Izumi et al. 2017, in preparation);\\
CAN: candidate of star-forming regions newly identified with the {\it WISE} data (see Section\ref{sec:3.1}); \\
others: ID of star-forming regions in the literature \citep{Snell02,Kobayashi08, Yasui08, Izumi14}.}
\tablenotetext{b}{A: aggregate, EC: embedded cluster, EC w/ H{\tiny II}: embedded cluster with H{\tiny II} region \citep{Snell02}}
\end{deluxetable*}
\begin{deluxetable*}{ccccccc}
\tabletypesize{\tiny}
\tablewidth{0pt}
\tablecaption{Contamination Rate for Molecular Clouds with Associating Sample Star-forming Regions \label{tbl:sampleContami}}
\tablehead{Molecular   & Number of Candidate/Sample  & Cloud          &Number Density of                   &Number of    &Number Density of               &Contamination\\
                     & Star-forming Regions             &  Area            & Candidate/Sample Star-forming     &Candidates     & Candidate Star-forming                      &\\
Clouds          & in the Cloud                          &                    &  Regions in the Cloud                  &in the Field      & Regions in the Field              &  Rate   \\
                     &($N_{\rm MC}$)                   &(arcmin$^2$)  & ($n_{\rm MC}$) (arcmin$^{-2}$) &($N_{\rm F}$)& ($n_{\rm F}$) (arcmin$^{-2}$)              &(\%) }
\startdata                   
Digel Cloud 1 (NRO data)      & 8                          & 24               &  3.3 $\times$ $10^{-1}$              &36                 & 1.3 $\times$ $10^{-2}$  & 4    \\
Digel Cloud 1 (FCRAO data)  & 8                          & 39                &  2.1 $\times$ $10^{-1}$              &36                 & 1.3 $\times$ $10^{-2}$  & 6    \\
Digel Cloud 2 (NRO data)      &25                         & 46                & 5.4 $\times$ $10^{-1}$               &40                 & 1.5 $\times$ $10^{-2}$  & 3    \\
Digel Cloud 2 (FCRAO data)  &25                         & 56                 & 4.5 $\times$ $10^{-1}$               &40                 & 1.5 $\times$ $10^{-2}$  & 3    \\
WB89-789 cloud     &\nodata                                &\nodata           &\nodata                                      &\nodata          &\nodata                          &\nodata   \\ 
01537+6154 cloud     &10                                       &130                & 7.7 $\times$ $10^{-2}$               &21                 &7.9 $\times$ $10^{-3}$   &10  \\
01587+6148 cloud     & 4                                        & 36                 & 1.1 $\times$ $10^{-1}$               &23                & 8.3 $\times$ $10^{-3}$  & 7  \\
02071+6235 cloud     & 1                                        & 41                &  2.4 $\times$ $10^{-2}$              & 30                & 7.2 $\times$ $10^{-3}$  & 29  \\       
02376+6030 cloud     & 6                                        & 72                & 8.3 $\times$ $10^{-2}$               &122               &4.5 $\times$ $10^{-2}$   & 54 \\
02383+6241 cloud     &12                                       & 28                & 4.3 $\times$ $10^{-1}$               &42                 &1.5 $\times$ $10^{-2}$   & 4  \\
02395+6244  cloud    & 0                                        & 21                &\nodata                                       &34                 &1.2 $\times$ $10^{-2}$   & \nodata  \\
02407+6029 cloud    &19                                       &130                & 1.5 $\times$ $10^{-1}$              &143                &5.4 $\times$ $10^{-2}$   & 37  \\
02413+6037 cloud     & 7                                        & 32                & 2.2 $\times$ $10^{-1}$               &147               &5.3 $\times$ $10^{-2}$   & 24  \\
02421+6233 cloud     & 2                                        & 18                & 1.1 $\times$ $10^{-1}$               &52                 &1.9 $\times$ $10^{-2}$   & 17  \\
02598+6008 cloud     & 4                                        & 28                &  1.4 $\times$ $10^{-1}$              &75                 & 2.7 $\times$ $10^{-2}$  & 19  \\                           
\enddata
\end{deluxetable*}



\appendix
\section{List of newly identified star-forming regions and their parental clouds}
Detailed information on newly identified star-forming regions and their parental molecular clouds is summrized in this appendix.
Table \ref{tbl:NewcanSFC} is for the star-forming molecular clouds, and Table \ref{tbl:NewcanSFR} is for the star-forming regions.

\begin{deluxetable}{lccccccccc}
\tabletypesize{\scriptsize}
\tablewidth{0pt}
\tablecaption{Molecular Clouds with Associated New Candidate Star-forming Regions \label{tbl:NewcanSFC}}
\tablehead{
Molecular Cloud &\multicolumn{2}{c}{Galactic Coordinate}&$V_{\rm LSR}$&$D$&$R_{\rm G}$&Number of Candidates&Contamination&Note$^a$\\
&$l$ (deg) &$b$ (deg) &(km s$^{-1}$)&(kpc)&(kpc)&in the Cloud&Rate (\%)& }
\startdata
$[$BKP2003$]$8672 & 139.344 & 2.891 & -52.87 & 5.85 & 13.5 & 4 & 33.0 & \\
$[$BKP2003$]$8674 & 139.397 & 2.768 & -53.34 & 5.93 & 13.6 & 4 & 12.9 & \\
$[$BKP2003$]$8713 & 136.314 & 0.165 & -57.64 & 6.23 & 13.7 & 15 & 21.9 & \\
$[$BKP2003$]$8715 & 136.269 & 0.260 & -56.09 & 5.98 & 13.5 & 1 & 231 & \\
$[$BKP2003$]$8911 & 139.096 & 3.518 & -54.32 & 6.06 & 13.7 & 5 & 6.72 & \\
$[$BKP2003$]$8918 & 141.464 & 4.484 & -50.66 & 5.80 & 13.5 & 2 & 18.0 & \\
$[$BKP2003$]$9254 & 136.789 & 0.445 & -59.82 & 6.65 & 14.1 & 11 & 31.5 & \\
$[$BKP2003$]$9255 & 136.866 & 0.257 & -56.94 & 6.18 & 13.7 & 3 & 105 & \\
\enddata
\tablenotetext{a}{Name of the 12 molecular clouds, which were used as molecular clouds with associated sample star-forming regions (Table \ref{tbl:sampleSFR}, Figures \ref{CO-NIR-MIR-DC12} and \ref{MIR-SnellSFRs}).
The CO peak of Digel Cloud 1a was not detected from the FCRAO CO survey of the outer Galaxy.}
\end{deluxetable}
\clearpage
\begin{deluxetable}{lccccccccccc}
\tabletypesize{\scriptsize}
\tablewidth{0pt}
\tablecaption{New Candidate Star-forming Regions\label{tbl:NewcanSFR}}
\tablehead{
Molecular Cloud & AllWISE& \multicolumn{2}{c}{Coordinate} & 3.4 $\mu$m   &$\sigma_{3.4}$   &4.6 $\mu$m   &$\sigma_{4.6}$& 12 $\mu$m    &$\sigma_{12}$    &22 $\mu$m    &$\sigma_{22}$ \\
& Source$^a$ &$l$ (deg)&$b$ (deg)&
(mag)                &(mag)                  &(mag)              &(mag)        &
(mag)               &(mag)                 &(mag)               &(mag)          }
\startdata
$[$BKP2003$]$8672 & J031418.89+610923.1 & 139.312 &  2.914 & 14.247 & 0.028 & 12.973 & 0.029 & 10.343 & 0.082 & 7.576 & 0.161 \\ 
                                & J031423.41+610435.3 & 139.361 &  2.850 & 13.967 & 0.030 & 13.165 & 0.029 & 10.388 & 0.076 & 7.811 & 0.166 \\ 
                                & J031428.58+610937.6 & 139.327 &  2.927 & 15.809 & 0.054 & 15.303 & 0.091 & 11.113 & 0.149 & 8.391 & null \\ 
                                & J031511.68+610614.5 & 139.430 &  2.925 & 16.654 & 0.088 & 14.935 & 0.068 & 11.607 & 0.194 & 8.579 & null \\ 
$[$BKP2003$]$8674 & J031351.58+610238.0 & 139.324 &  2.789 & 12.832 & 0.024 & 11.965 & 0.023 & 9.549 & 0.058 & 6.803 & 0.083 \\ 
                                & J031407.57+605939.0 & 139.377 &  2.763 & 14.185 & 0.028 & 12.837 & 0.026 & 10.373 & 0.080 & 8.443 & 0.288 \\ 
                                & J031415.70+605652.0 & 139.415 &  2.732 & 14.321 & 0.030 & 13.754 & 0.037 & 10.811 & 0.093 & 8.128 & 0.237 \\ 
                                & J031416.94+605737.4 & 139.411 &  2.744 & 14.737 & 0.039 & 13.455 & 0.034 & 10.667 & 0.087 & 7.642 & 0.154 \\
\enddata
\tablenotetext{a}{More than one cloud at different velocities exists in the line of sight for the following 13 candidates:
J024735.40+594927.3, J024427.51+604136.2, J024440.19+604230.6, J005521.72+642642.8, J005538.36+642725.9,
J024401.92+624822.3, J024129.21+604327.8, J231310.32+630839.3, J231317.89+631031.3, J225340.98+623331.4,
J000625.70+635804.3, J000626.43+640035.1, and J000636.77+640033.3.}
\end{deluxetable}

\begin{thebibliography}{}
\bibitem[Anderson et al.(2014)]{Anderson14} Anderson, L.~D., Bania, T.~M., Balser, D.~S., et al.\ 2014, \apjs, 212, 1
\bibitem[Anderson et al.(2015)]{Anderson15} Anderson, L.~D., Armentrout, W.~P., Johnstone, B.~M., et al.\ 2015, \apjs, 221, 26  
\bibitem[Bastian et al.(2010)]{Bastian10} Bastian, N., Covey, K.~R., \& Meyer, M.~R.\ 2010, \araa, 48, 339
\bibitem[Beichman et al.(1988)]{Beichman88} Beichman, C.~A., Neugebauer, G., Habing, H.~J., Clegg, P.~E., \& Chester, T.~J.\ 1988, Infrared astronomical satellite (IRAS) catalogs and atlases.~Volume 1: Explanatory supplement, 1,  
\bibitem[Bigiel et al.(2008)]{Bigiel08} Bigiel, F., Leroy, A., Walter, F., et al.\ 2008, \aj, 136, 2846 
\bibitem[Bolatto et al.(2008)]{Bolatto08} Bolatto, A.~D., Leroy, A.~K., Rosolowsky, E., Walter, F., \& Blitz, L.\ 2008, \apj, 686, 948 
\bibitem[Bolatto et al.(2013)]{Bolatto13} Bolatto, A.~D., Wolfire, M., \& Leroy, A.~K.\ 2013, \araa, 51, 207
\bibitem[Brand \& Wouterloot(1994)]{Brand94} Brand, J., \& Wouterloot, J.~G.~A.\ 1994, \aaps, 103, 503
\bibitem[Brand \& Wouterloot(1995)]{Brand95} Brand, J., \& Wouterloot, J.~G.~A.\ 1995, \aap, 303, 851 
\bibitem[Brand \& Wouterloot(2007)]{Brand07} Brand, J., \& Wouterloot, J.~G.~A.\ 2007, \aap, 464, 909 
\bibitem[Brunt et al.(2000)]{Brunt00} Brunt, C.~M., Ontkean, J., \& Knee, L.~B.~G.\ 2000, Bulletin of the American Astronomical Society, 197, 508 
\bibitem[Brunt et al.(2003)]{Brunt03} Brunt, C.~M., Kerton, C.~R., \& Pomerleau, C.\ 2003, \apjs, 144, 47  
\bibitem[Cox(2000)]{Cox00} Cox, A.~N.\ 2000, Allen's Astrophysical Quantities,  
\bibitem[da Cunha et al.(2008)]{Cunha08} da Cunha, E., Charlot, S., \& Elbaz, D.\ 2008, \mnras, 388, 1595 
\bibitem[Dame et al.(2001)]{Dame01} Dame, T.~M., Hartmann, D., \& Thaddeus, P.\ 2001, \apj, 547, 792 
\bibitem[Digel et al.(1994)]{Digel94} Digel, S., de Geus, E., \& Thaddeus, P.\ 1994, \apj, 422, 92
\bibitem[Elia et al.(2013)]{Elia13} Elia, D., Molinari, S., Fukui, Y., et al.\ 2013, \apj, 772, 45 
\bibitem[Ferguson et al.(1998)]{Ferguson98} Ferguson, A.~M.~N., Gallagher, J.~S., \& Wyse, R.~F.~G.\ 1998, \aj, 116, 673 
\bibitem[Frew \& Parker(2006)]{Frew06} Frew, D.~J., \& Parker, Q.~A.\ 2006, Planetary Nebulae in our Galaxy and Beyond, 234, 49
\bibitem[Frew(2008)]{Frew08} Frew, D.~J.\ 2008, Ph.D.~Thesis,  
\bibitem[Georgelin \& Georgelin(1976)]{Georgelin76} Georgelin, Y.~M., \& Georgelin, Y.~P.\ 1976, \aap, 49, 57 
\bibitem[Hachisuka et al.(2015)]{Hachisuka15} Hachisuka, K., Choi, Y.~K., Reid, M.~J., et al.\ 2015, \apj, 800, 2 
\bibitem[Heyer et al.(1998)]{Heyer98} Heyer, M.~H., Brunt, C., Snell, R.~L., et al.\ 1998, \apjs, 115, 241
\bibitem[Heyer et al.(2001)]{Heyer01} Heyer, M.~H., Carpenter, J.~M., \& Snell, R.~L.\ 2001, \apj, 551, 852 
\bibitem[Hou \& Han(2014)]{Hou14} Hou, L.~G., \& Han, J.~L.\ 2014, \aap, 569, A125 
\bibitem[Hughes \& MacLeod(1989)]{Hughes89} Hughes, V.~A., \& MacLeod, G.~C.\ 1989, \aj, 97, 786 
\bibitem[Izumi et al.(2014)]{Izumi14} Izumi, N., Kobayashi, N., Yasui, C., et al.\ 2014, \apj, 795, 66 
\bibitem[Jarrett et al.(2011)]{Jarrett11} Jarrett, T.~H., Cohen, M., Masci, F., et al.\ 2011, \apj, 735, 112 
\bibitem[Kalberla et al.(2005)]{Kalberla05} Kalberla, P.~M.~W., Burton, W.~B., Hartmann, D., et al.\ 2005, \aap, 440, 775 
\bibitem[Kennicutt(1998)]{Kennicutt98} Kennicutt, R.~C., Jr.\ 1998, \apj, 498, 541 
\bibitem[Kennicutt \& Evans(2012)]{Kennicutt12} Kennicutt, R.~C., \& Evans, N.~J.\ 2012, \araa, 50, 531 
\bibitem[Kerton \& Brunt(2003)]{Kerton03} Kerton, C.~R., \& Brunt, C.~M.\ 2003, \aap, 399, 1083
\bibitem[Kobayashi \& Tokunaga(2000)]{Kobayashi00} Kobayashi, N., \& Tokunaga, A.~T.\ 2000, \apj, 532, 423
\bibitem[Kobayashi et al.(2008)]{Kobayashi08} Kobayashi, N., Yasui, C., Tokunaga, A.~T., \& Saito, M.\ 2008, \apj, 683, 178 
\bibitem[Koenig \& Leisawitz(2014)]{Koenig14} Koenig, X.~P., \& Leisawitz, D.~T.\ 2014, \apj, 791, 131 
\bibitem[Kutner(1983)]{Kutner83} Kutner, M.~L.\ 1983, Surveys of the Southern Galaxy, 105, 143
\bibitem[Lada \& Lada(2003)]{Lada03} Lada, C.~J., \& Lada, E.~A.\ 2003, \araa, 41, 57
\bibitem[McClure(2009)]{McClure09} McClure, M.\ 2009, \apjl, 693, L81 
\bibitem[May et al.(1993)]{May93} May, J., Bronfman, L., Alvarez, H., Murphy, D.~C., \& Thaddeus, P.\ 1993, \aaps, 99, 105 
\bibitem[May et al.(1997)]{May97} May, J., Alvarez, H., \& Bronfman, L.\ 1997, \aap, 327, 325 
\bibitem[Mead \& Kutner(1988)]{Mead88} Mead, K.~N., \& Kutner, M.~L.\ 1988, \apj, 330, 399 
\bibitem[Moe \& De Marco(2006)]{Moe06} Moe, M., \& De Marco, O.\ 2006, \apj, 650, 916 
\bibitem[Nakanishi \& Sofue(2003)]{Nakanishi03} Nakanishi, H., \& Sofue, Y.\ 2003, \pasj, 55, 191 
\bibitem[Nakanishi \& Sofue(2006)]{Nakanishi06} Nakanishi, H., \& Sofue, Y.\ 2006, \pasj, 58, 847 
\bibitem[Nakagawa et al.(2005)]{Nakagawa05} Nakagawa, M., Onishi, T., Mizuno, A., \& Fukui, Y.\ 2005, \pasj, 57, 917
\bibitem[Neugebauer et al.(1984)]{Neugebauer84} Neugebauer, G., Habing, H.~J., van Duinen, R., et al.\ 1984, \apjl, 278, L1 
\bibitem[Reid et al.(2014)]{Reid14} Reid, M.~J., Menten, K.~M., Brunthaler, A., et al.\ 2014, \apj, 783, 130 
\bibitem[Rice et al.(2016)]{Rice16} Rice, T.~S., Goodman, A.~A., Bergin, E.~A., Beaumont, C., \& Dame, T.~M.\ 2016, \apj, 822, 52
\bibitem[Rosolowsky(2005)]{Rosolowsky05} Rosolowsky, E.\ 2005, \pasp, 117, 1403 
\bibitem[Rudolph et al.(1996)]{Rudolph96} Rudolph, A.~L., Brand, J., de Geus, E.~J., \& Wouterloot, J.~G.~A.\ 1996, \apj, 458, 653 
\bibitem[Santos et al.(2000)]{Santos00} Santos, C.~A., Yun, J.~L., Clemens, D.~P., \& Agostinho, R.~J.\ 2000, \apjl, 540, L87 
\bibitem[Schruba et al.(2011)]{Schruba11} Schruba, A., Leroy, A.~K., Walter, F., et al.\ 2011, \aj, 142, 37 
\bibitem[Shi et al.(2014)]{Shi14} Shi, Y., Armus, L., Helou, G., et al.\ 2014, \nat, 514, 335 
\bibitem[Schmidt(1959)]{Schmidt59} Schmidt, M.\ 1959, \apj, 129, 243 
\bibitem[Smartt et al.(1996)]{Smartt96} Smartt, S.~J., Dufton, P.~L., \& Rolleston, W.~R.~J.\ 1996, \aap, 305, 164 
\bibitem[Smartt \& Rolleston(1997)]{Smartt97} Smartt, S.~J., \& Rolleston, W.~R.~J.\ 1997, \apjl, 481, L47 
\bibitem[Snell et al.(2002)]{Snell02} Snell, R.~L., Carpenter, J.~M., \& Heyer, M.~H.\ 2002, \apj, 578, 229
\bibitem[Strasser et al.(2007)]{Strasser07} Strasser, S.~T., Dickey, J.~M., Taylor, A.~R., et al.\ 2007, \aj, 134, 2252 
\bibitem[Sun et al.(2015)]{Sun15} Sun, Y., Xu, Y., Yang, J., et al.\ 2015, \apjl, 798, L27 
\bibitem[Taylor et al.(2003)]{Taylor03} Taylor, A.~R., Gibson, S.~J., Peracaula, M., et al.\ 2003, \aj, 125, 3145 
\bibitem[Thilker et al.(2005)]{Thilker05} Thilker, D.~A., Bianchi, L., Boissier, S., et al.\ 2005, \apjl, 619, L79
\bibitem[Vall{\'e}e(2008)]{Vallee08} Vall{\'e}e, J.~P.\ 2008, \aj, 135, 1301
\bibitem[V{\'a}zquez et al.(2008)]{Vazquez08} V{\'a}zquez, R.~A., May, J., Carraro, G., et al.\ 2008, \apj, 672, 930-939 
\bibitem[Werner et al.(2004)]{Werner04} Werner, M.~W., Roellig, T.~L., Low, F.~J., et al.\ 2004, \apjs, 154, 1 
\bibitem[Wolfire et al.(2003)]{Wolfire03} Wolfire, M.~G., McKee, C.~F., Hollenbach, D., \& Tielens, A.~G.~G.~M.\ 2003, \apj, 587, 278 
\bibitem[Wouterloot \& Brand(1989)]{Wouterloot89} Wouterloot, J.~G.~A., \& Brand, J.\ 1989, \aaps, 80, 149 
\bibitem[Wright et al.(2010)]{Wright10} Wright, E.~L., Eisenhardt, P.~R.~M., Mainzer, A.~K., et al.\ 2010, \aj, 140, 1868 
\bibitem[Yang et al.(2002)]{Yang02} Yang, J., Jiang, Z., Wang, M., Ju, B., \& Wang, H.\ 2002, \apjs, 141, 157 
\bibitem[Yasui et al.(2006)]{Yasui06} Yasui, C., Kobayashi, N., Tokunaga, A.~T., Terada, H., \& Saito, M.\ 2006, \apj, 649, 753
\bibitem[Yasui et al.(2008)]{Yasui08} Yasui, C., Kobayashi, N., Tokunaga, A.~T., Terada, H., \& Saito, M.\ 2008, \apj, 675, 443
\bibitem[Yun et al.(2015)]{Yun15} Yun, J.~L., Elia, D., Djupvik, A.~A., Torrelles, J.~M., \& Molinari, S.\ 2015, \mnras, 452, 1523  
\end{thebibliography}
\end{document}